\documentclass[a4paper, preprint, aps, prb, amsmath, amssymb]{revtex4-1}

\usepackage[utf8]{inputenc}
\usepackage{graphicx}
\usepackage{bm}

\newcommand{\mus}{\ensuremath{\mu \mathrm{s}}}

\newcommand{\Ap}{\ensuremath{A_\mathrm{p}}}
\newcommand{\mi}{\ensuremath{m_\mathrm{i}}}

\newcommand{\Is}{\ensuremath{I_\mathrm{s}}}
\newcommand{\nee}{\ensuremath{n_\mathrm{e}}}
\newcommand{\Te}{\ensuremath{T_\mathrm{e}}}
\newcommand{\Vf}{\ensuremath{V_\mathrm{f}}}
\newcommand{\Vp}{\ensuremath{V_\mathrm{p}}}
\newcommand{\Vrad}{\ensuremath{U}}
\newcommand{\Gamman}{\ensuremath{\Gamma_{n}}}
\newcommand{\Gammat}{\ensuremath{\Gamma_{T}}}

\newcommand{\Istilde}{\ensuremath{\widetilde{I}_\mathrm{s}}}
\newcommand{\netilde}{\ensuremath{\widetilde{n}_\mathrm{e}}}
\newcommand{\Tetilde}{\ensuremath{\widetilde{T}_\mathrm{e}}}
\newcommand{\Vftilde}{\ensuremath{\widetilde{V}_\mathrm{f}}}
\newcommand{\Vptilde}{\ensuremath{\widetilde{V}_\mathrm{p}}}
\newcommand{\Utilde}{\ensuremath{\widetilde{U}}}
\newcommand{\Gammanhat}{\ensuremath{\widehat{\Gamma}_\mathrm{n}}}
\newcommand{\Gammathat}{\ensuremath{\widehat{\Gamma}_\mathrm{T}}}

\newcommand{\mean}[1]{\ensuremath{\langle #1 \rangle}}
\newcommand{\ma}[1]{\ensuremath{{\langle {#1} \rangle_\mathrm{mv}}}}
\newcommand{\mrms}[1]{\ensuremath{{{#1}_\mathrm{rms,mv}}}}
\newcommand{\taud}{\ensuremath{\tau_\mathrm{d}}}
\newcommand{\tauw}{\ensuremath{\tau_\mathrm{w}}}
\newcommand{\tauf}{\ensuremath{\tau_\mathrm{f}}}
\newcommand{\taur}{\ensuremath{\tau_\mathrm{r}}}

\newcommand{\Ref}[1]{[\onlinecite{#1}]}
\newcommand{\Eqnref}[1]{Eq. (~\ref{eq:#1})}
\newcommand{\Eqsref}[1]{Eqs. (~\ref{eq:#1})}
\newcommand{\Figref}[1]{Fig.~\ref{fig:#1}}
\newcommand{\Figsref}[1]{Figs.~\ref{fig:#1}}
\newcommand{\Tabref}[1]{Tab.~\ref{tab:#1}}
\newcommand{\Secref}[1]{Sec.~\ref{sec:#1}}
\newcommand{\Secsref}[1]{Secs.~\ref{sec:#1}}

\begin{document}
\title{Intermittent electron density and temperature fluctuations and associated fluxes in the Alcator C-Mod scrape-off layer}
\author{R.~Kube}
\email[E-mail:]{ralph.kube@uit.no}
\author{O.~E.~Garcia}
\author{A.~Theodorsen}
\affiliation{Department of Physics and Technology, UiT The Arctic University of Norway, N-9037 Tromsø, Norway}
\author{D.~Brunner}
\author{A.~Q. Kuang}
\author{B.~LaBombard}
\author{J.~L.~Terry}
\affiliation{MIT Plasma Science and Fusion Center, Cambridge, MA, 02139, USA}

\date{\today}
\begin{abstract}
    The Alcator C-Mod mirror Langmuir probe system has been used to sample data time
    series of fluctuating plasma parameters in the outboard mid-plane far scrape-off layer.
    We present a statistical analysis
    of one second long time series of electron density, temperature, radial electric drift
    velocity and the corresponding particle and electron heat fluxes. These are sampled
    during stationary plasma conditions in an ohmically heated, lower single null diverted 
    discharge.
    The electron density and temperature are strongly correlated and feature
    fluctuation statistics similar to the ion saturation current. Both electron density and
    temperature time series are dominated by intermittent, large-amplitude burst
    with an exponential distribution of both burst amplitudes and waiting times between
    them.
    The characteristic time scale of the large-amplitude bursts is approximately
    $15\, \mus$. Large-amplitude velocity fluctuations feature a slightly faster
    characteristic time scale and appear at a faster rate than electron density and 
    temperature fluctuations. 
    Describing these time series as a superposition of uncorrelated exponential
    pulses, we find that probability distribution functions, power spectral
    densities as well as auto-correlation functions of the data time series
    agree well with predictions from the stochastic model. 
    The electron particle and heat fluxes present large-amplitude fluctuations.
    For this low-density plasma, the radial electron heat flux is dominated by
    convection, that is, correlations of fluctuations in the electron density
    and radial velocity. Hot and dense blobs contribute approximately 
    $6\%$ of the total fluctuation driven heat flux.
\end{abstract}

\maketitle

\section{Introduction}
% Why scrape-off layer plasma is interesting
Turbulent flows in the scrape-off layer (SOL) of magnetically confined plasmas have received 
great attention recently. Experimental analyses have demonstrated that plasma blobs propagating 
through the scrape-off layer towards the vessel wall dominate the plasma particle and heat fluxes 
at the outboard mid-plane
\Ref{labombard-2001, rudakov-2002, boedo-2003, rudakov-2005, garcia-2007-nf, garcia-2007-nf, 
     garcia-2008-ppcf, fedorczak-2009, carralero-2014, graves-2005, antar-2001, walkden-2017}.
In order to assess expected erosion and damage to the plasma enclosing vessel,
the statistics of the impinging plasma fluxes are of great interest
\Ref{federici-2001, loarte-2007_nf, lipschultz-2007, wenninger-2017}. 

Plasma blobs are pressure perturbations spatially localized in the plane 
perpendicular to the magnetic field and elongated along the magnetic field lines.
They are believed to be created in the vicinity of the last closed magnetic flux surface
with particle density perturbation amplitudes comparable in magnitude to the average
scrape-off layer particle density. At the outboard mid-plane location the magnetic
curvature vector and field strength gradient point towards the magnetic axis.
This causes an electric polarization of the blob structure due to magnetic
curvature and gradient drifts. The resulting electric field propagates 
the blob towards the vessel wall, resulting in large cross-field particle
and heat fluxes onto plasma facing components.
\Ref{krash-2001, dippolito-2002, bian-2003, yu-2003, myra-2004, aydemir-2005, garcia-2006}.

Scrape-off layer plasma fluctuations furthermore exhibit several universal features.  Time
series data of plasma density fluctuations feature non-gaussian values of sample skewness and 
flatness and their probability density functions (PDFs)
present elevated tails for large amplitude events. This feature has been
observed in experiments
\Ref{antar-2001, antar-2003, graves-2005, xu-2005, garcia-2013, kube-2016-ppcf, 
     theodorsen-2016-php, theodorsen-2017-nf, garcia-2016-nme, walkden-2017}
as well as in numerical simulations
\Ref{ghendrih-2003, garcia-2004, bisai-2005, garcia-2008-ppcf, militello-2012}
and is well documented to be due to the radial propagation of plasma blobs
\Ref{terry-2003, zweben-2004, terry-2005, zweben-2006, agostini-2011, maqueda-2011, 
    garcia-2013, kube-2013}.
A quadratic relation between sample skewness and flatness has been reported
from several experiments 
\Ref{labit-2007, sattin-2009, banerjee-2012, garcia-2013-jnm, banerjee-2014, kube-2016-ppcf, theodorsen-2016-ppcf, garcia-2016}.
Conditionally averaged waveforms of electron density time series exhibit
approximately two-sided exponential waveform shapes
\Ref{boedo-2003, rudakov-2005, garcia-2007-nf, silva-2009, garcia-2013, 
     tanaka-2015, theodorsen-2016-ppcf, garcia-2016, kube-2016-ppcf}.
Several experiments report large-amplitude electron density fluctuations
in phase with an outwards $\bm{E} \times \bm{B}$ drift velocity, that is radial particle
flux events
\Ref{boedo-2003, grulke-2006, saha-2008, silva-2009, kube-2016-ppcf, theodorsen-2016-ppcf}.

A recently developed stochastic model describes such time series as a super-position of
uncorrelated pulses \Ref{garcia-2012}. Assuming an exponential pulse shape and exponentially
distributed pulse amplitudes and waiting times between pulses
\Ref{garcia-2013, garcia-2013-jnm, garcia-2015, theodorsen-2016-php, garcia-2016-nme}
it predicts the fluctuation amplitudes to be Gamma distributed. The quadratic relation
between moments of skewness and flatness of a Gamma distributed variable is in excellent
agreement with the quadratic relation observed in experiments. This model furthermore 
predicts the experimentally observed exponential density profiles in scrape-off layer plasmas 
\Ref{garcia-2016, militello-2016}.
The stochastic model has been generalized to describe general pulse shapes as well as
additive noise. Analytic expressions of probability density functions, auto-correlation
functions, power spectral densities and level crossing rates have been derived
\Ref{theodorsen-2016-php, garcia-2016, theodorsen-2017-ps, garcia-2017-ac}. In this
contribution it is demonstrated that the model predictions compare favorably with measurements
of the fluctuating electron density and temperature, as well as with the radial velocity. 
It should be noted that by constructing the stochastic model as a superposition of individual
pulses, the underlying non-linear dynamics of the plasma is parameterized. Specifically,
the steepening of radially propagating blob structures is modeled by exponential pulse 
shapes. Another approach, which proposes a stochastic differential equations to describe 
the non-linear plasma dynamics, under the constraint that the fluctuations are Gamma
distributed, has recently been explored \Ref{mekkaoui-2013}.

% What is used to investigate the scrape-off layer
Scrape-off layer plasmas are usually diagnosed with Langmuir probes. They allow for three
fundamental modes of operation. One way is to apply a sweeping voltage to a Langmuir 
electrode. This allows the plasma density, the electric potential as well as the
electron temperature to be inferred on the time scale of the sweeping voltage. This time
scale is commonly of the orders of milliseconds, as to avoid hysteresis effects which arise
at higher sweeping frequencies \Ref{muller-2010}. On the other hand, the time scale
associated with blob propagation is on the order of microseconds. Conventional sweeping
modes can thus not be used to investigate plasma fluctuations.

A second way is to bias the Langmuir electrode to a large negative electric potential
relative to the vacuum vessel. This way the electrode draws the ion saturation current
\Ref{hutchinson-book}
\begin{align}
    \Is & = \frac{1}{2} e \nee \Ap \sqrt{\frac{k_\mathrm{b} \Te}{\mi}}. \label{eq:Isat}
\end{align}
Here $e$ is the elementary charge, 
$\nee$ is the electron density,
$\Te$ is the electron temperature,
$\mi$ is the ion mass and 
$A_\mathrm{p}$ is the current collecting probe area.
This assumes that the ion temperature is zero, although it is typically larger than the
electron temperature for the measurements in this paper. Employing a Reynolds decomposition
of the time-dependent quantities in \Eqnref{Isat}, as $u(t) = u_0 + \widetilde{u}(t)$ for a
variable u, shows that fluctuations in $\nee$ and $\Te$ perturb the ion saturation current
as
\begin{align}
    \frac{\Istilde}{I_{\mathrm{s}, 0}} \approx \left( \frac{\netilde}{n_{\mathrm{e}, 0}} + \frac{1}{2} \frac{\Tetilde}{T_{\mathrm{e}, 0}} \right).
\end{align}
Here the factor $1/2$ comes from an expansion of the square-root for small relative electron
temperature fluctuations.
%
% Compute part due to density fluctuations: Isat^*(delta ne) / Isat 
% and relative part due to temperature fluctuations: Isat^*(delta Te) / Isat
%
For equal relative fluctuations of the electron density and temperature the
electron density contributes twice as much to the relative fluctuation of the ion
saturation current than the electron temperature fluctuation. With no fast measurements
of $\Te$ at hand, a constant value is often assumed for $\Te$ in \Eqnref{Isat} to 
find $\nee$ given $\Is$.

A third mode of operation is to electrically isolate the Langmuir electrode. In this mode,
it assumes the floating potential 
\begin{align}
    \Vf & = \Vp - \Lambda \Te,
\end{align}
where $\Vp$ is the plasma potential and $\Lambda \approx 2-3$ in scrape-off layer plasmas
\Ref{stangeby-book, rohde-1996}. Using again a Reynolds decomposition, the fluctuating 
floating potential is given by $\Vftilde = \Vptilde - \Lambda \Tetilde$. Thus, perturbations in the
floating potential are equally due to fluctuations in the plasma potential and the electron
temperature. 
Fast measurements of $\Te$ in the scrape-off layer are often unavailable such that 
$\nee$ is approximated by $\Is$ and the plasma potential is estimated by the floating 
potential. From this, an estimate of the radial $\bm{E} \times \bm{B}$ drift velocity can 
be calculated given two spatially separated measurements of the floating potential. It was 
recently observed that perturbations of the electron temperature may alter the estimated 
radial drift velocity 
\Ref{horacek-2010, nold-2012, gennrich-2012}.

Since plasma blobs present perturbations of the plasma density, temperature and electric
potential, real time measurements of all three quantities from a single point are desirable
as to precisely quantify their contributions to cross-field transport in the scrape-off layer.
Recent probe designs, such as ball pen probes \Ref{adamek-2004, adamek-2009} and emissive probes
\Ref{schrittwieser-2002, schrittwieser-2008, ionita-2011, adamek-2014} allow fast sampling
of the plasma potential but to evaluate the electron temperature one still needs to combine
data from multiple electrodes.

Langmuir probe implementations that utilize multiple electrodes to provide real time samples
of the fluctuating plasma parameters, such as triple probes, are routinely operated in
several major tokamaks \Ref{lin-1992, asakura-1995, yang-2003, kim-2016}.
In this configuration current and voltage samples from different Langmuir electrodes are
combined as to estimate the fluctuating electron density, the plasma potential, and
the electron temperature in real time. On the other hand the equations of the triple probe
configuration assume that the electrodes sample a homogeneous plasma. These assumptions are
often violated in the scrape-off layer, where the characteristic length of the turbulence
structures may be smaller than the separation of the Langmuir electrodes. 
Triple probe configurations have also been implemented in the time domain, removing the
assumption of a homogeneous plasma \Ref{meier-1995, meier-1997, meier-2001}. This
configuration requires two spatially separated Langmuir electrodes. Periodically biasing
the electrodes to three different bias voltages allows to infer the electron density and 
temperature, as well as the plasma potential at each Langmuir electrode independently.

Fast measurements of electron temperature fluctuation in scrape-off layer plasmas are sparse.
Measurements based on the method of harmonics \Ref{boedo-1999} taken in the DIII-D tokamak suggest
that fluctuations of the electron temperature and the electric drift velocity appear on average in
phase with fluctuations in the electron density \Ref{rudakov-2002, rudakov-2005}. However, the method
of harmonics has a time resolution of $10\, \mus$, comparable to the time scale of the turbulence
structures in the plasma \Ref{boedo-1999}. Analysis of an $8\, \mathrm{ms}$ long electron temperature
data time series, taken by a triple probe configuration in the SINP tokamak, suggests that it presents
the same non-gaussian features as commonly observed in electron density time series: the frequent arrival
of large-amplitude bursts and heavy-tailed histograms \Ref{saha-2008}. Recent measurements reported
from ASDEX Upgrade confirm that fluctuations of the electron density and temperature appear in phase,
together with fluctuations in the plasma potential \Ref{horacek-2010}. It was furthermore reported
that the temperature fluctuations show on average a temporal asymmetry around the density peaks.
Relative fluctuation levels of the electron temperature were found to be lower by a factor of
approximately $2-3$ than for the electron density.

The novel mirror Langmuir Probe (MLP) biasing technique allows for fast sampling of the ion saturation
current, the electron temperature and the floating potential at a single sampling position 
\Ref{labombard-2007, labombard-2014}. This diagnostics consists of three major 
components. The actual mirror Langmuir probe is an electronic circuit that generates 
a current-voltage (I-V) characteristic with the three adjustable parameters $\Is$, $\Te$, and $\Vf$:
\begin{align}
    I_\mathrm{MLP} & = \Is \left[ \exp\left( \frac{V - \Vf}{\Te} \right) - 1\right]
\end{align}
The second main component is a Langmuir electrode immersed in the plasma to be
sampled. Both components are connected to a fast switching biasing waveform.
The bias waveform switches between the states $(V^{+}, V^{0}, V^{-})$, such that the Langmuir
electrode draws approximately $\pm \Is$ at the states $V^{\mp}$ and zero net current when biased to
$V^{0}$, as shown in Fig.~1 of \Ref{labombard-2014}. Every $300\, \mathrm{ns}$ the bias voltage
state is updated. Current samples from the MLP and the Langmuir electrode are compared after the
bias voltage has settled. In order to minimize the deviation between the two sample pairs, the MLP
adjusts the $\Is$, $\Te$, and $\Vf$ parameters dynamically.
The main task of the MLP circuit is to set and maintain the optimal range of the bias voltages such
that a complete I-V characteristic can be reconstructed from measurements at the Langmuir electrode
at the three bias voltages states. Samples of the I-V response from the MLP and the Langmuir electrode
are digitized at $3\, \mathrm{MHz}$, synchronized to the states $V^{+}$, $V^{0}$, and $V^{-}$. The
current and voltage samples of the Langmuir electrode are then used for a fit to the I-V characteristic
as to obtain $\Te$, $\Vf$ and $\Is$. The time staggering of the three sequential measurements voltages
is neglected. Time series of the fit parameters at a sampling frequency of $1\, \mathrm{MHz}$ are
obtained by mapping them one-to-one to the time samples of the voltage states $V^{+}$, $V^{0}$ and
$V^{-}$. Finally, the data time series of the $\Te$, $\Vf$, and $\Is$ fit parameters are linearly
interpolated on the same time base with $3\, \mathrm{MHz}$ sampling frequency. From these sample
values the electron density and the plasma potential are calculated \Ref{labombard-2014}.

This contribution presents a statistical analysis of exceptionally long data time series measured
by the MLP in stationary plasma conditions. Section~\ref{sec:setup} describes the experimental setup
and \Secref{analysis} describes the data analysis methods employed. The statistical properties of the
ion saturation current, floating potential, as well as electron density and temperature are discussed
in \Secref{single}. Fluctuation time series of the radial velocity, the radial electron particle and
heat fluxes are analyzed in \Secref{multi}. A discussion and a conclusion of the results are given in
\Secsref{discussion} and~\ref{sec:conclusion}. Supplementary information on the stochastic model and
on analysis of the MLP data is given in the Appendices A and B.

%%%%%%%%%%%%%%%%%%%%%%%%%%%%%%%%%%%%%%%%%%%%%%%%%%%%%%%%%%%%%%%%%%%%%%%%%%%%%%%%%%%%%%%%%%%%%%%%%%%
\section{Experimental setup}
\label{sec:setup}
Alcator C-Mod is a compact, high-field tokamak with major radius $R = 0.68\, \mathrm{m}$ and minor
radius $a = 0.21\, \mathrm{m}$ \Ref{hutchinson-1994, greenwald-2013, greenwald-2014}. It allows for
an on-axis magnetic field strength of up to $8\, \mathrm{T}$ so as to confine plasmas with up
to two atmospheres pressure. In this contribution we investigate the outboard mid-plane scrape-off
layer of an ohmically heated plasma in a lower single-null diverted magnetic configuration with an 
on-axis toroidal field of $B_\mathrm{T} = 5.4\, \mathrm{T}$. The toroidal plasma current for the 
investigated discharge is $I_\mathrm{p} = 0.55\, \mathrm{MA}$ and the line averaged core plasma
density is given by $\overline{n}_\mathrm{e} / n_\mathrm{G} = 0.12$, where $n_\mathrm{G}$ is the
Greenwald density. Such low density plasmas feature a far scrape-off layer with vanishing electron
pressure gradients along magnetic field lines. The temperature drop from outboard mid-plane to the
divertor plates is supported by the divertor sheaths \Ref{lipschultz-2007-fst}.

A Mach probe head was dwelled at the limiter radius, approximately 
$0.11\, \mathrm{m}$ above the outboard mid-plane location. Its four electrodes are 
arranged in a pyramidal dome geometry on the probe head such that they sample approximately 
the same magnetic flux surface. Each electrode is connected to a MLP bias drive, and labeled northeast, 
southeast, southwest and northwest. Tracing a magnetic field line from the outboard 
mid-plane to the probe head, the east electrodes are in the shadow of the west electrodes,
with the south electrodes facing the outboard mid-plane. The MLPs obtain $\Is$, $\Vf$, 
and $\Te$ from fits to the I-V samples with a sampling rate of approximately
$1\, \mathrm{MHz}$. Further details on the probe head are given in \Ref{smick-2009}.

\section{Data analysis}
\label{sec:analysis}

MLPs have been used successfully to measure profiles of average values and relative fluctuation
levels \Ref{labombard-2014}. However, large-amplitude fluctuations in the far scrape-off layer
present challenges to interpreting the reported fit values $\Is$, $\Vf$, and $\Te$. The MLP dynamically updates the voltage states $V^{+}$ and $V^{-}$ 
relative to a running average of electron temperature samples over a ~$2\, \mathrm{ms}$ window,
$V^{+} - V^{-} < 4 \, \overline{T}_{\mathrm{e}}$ holds, where $\overline{T}_\mathrm{e}$ denotes
this running average. When the instantaneous electron
temperature at the Langmuir electrode significantly exceeds $\overline{T}_{\mathrm{e}}$, the
range of the biasing voltages may be insufficient to resolve the I-V characteristic. This
can result in large uncertainties of the fit parameters. Moreover, events unrelated to the
turbulent plasma flows, such as probe arcing, may also produce in erroneous values of the
fit values.

Parameters from I-V fits reported from all four MLPs at a given time were compared 
to investigate the robustness of the measured fluctuations. It was found that 
$\Is$, $\Te$, and $\Vf$ fit values reported from the four MLPs are of comparable magnitude when
$V^{+} - V^{-} > 4\, \overline{T}_{\mathrm{e}}$ holds. On the other hand
the four $\Te$ values may feature large outliers when
$V^{+} - V^{-} < 4\, \overline{T}_{\mathrm{e}}$ holds.
Therefore we analyze data time series obtained by applying a 12-point Gaussian filter on the current
time samples obtained at the electrode biasing potentials $(V^{+}, V^{0}, V^{-})$. The $\Is$, $\Vp$
and $\Vf$ time series used throughout this article are taken from the southwest MLP. The used $\nee$
and $\Te$ time series are given by the average of the fit parameters reported by all four MLPs. 

\begin{figure}
    % created by mkplot_time_traces.py
    \includegraphics{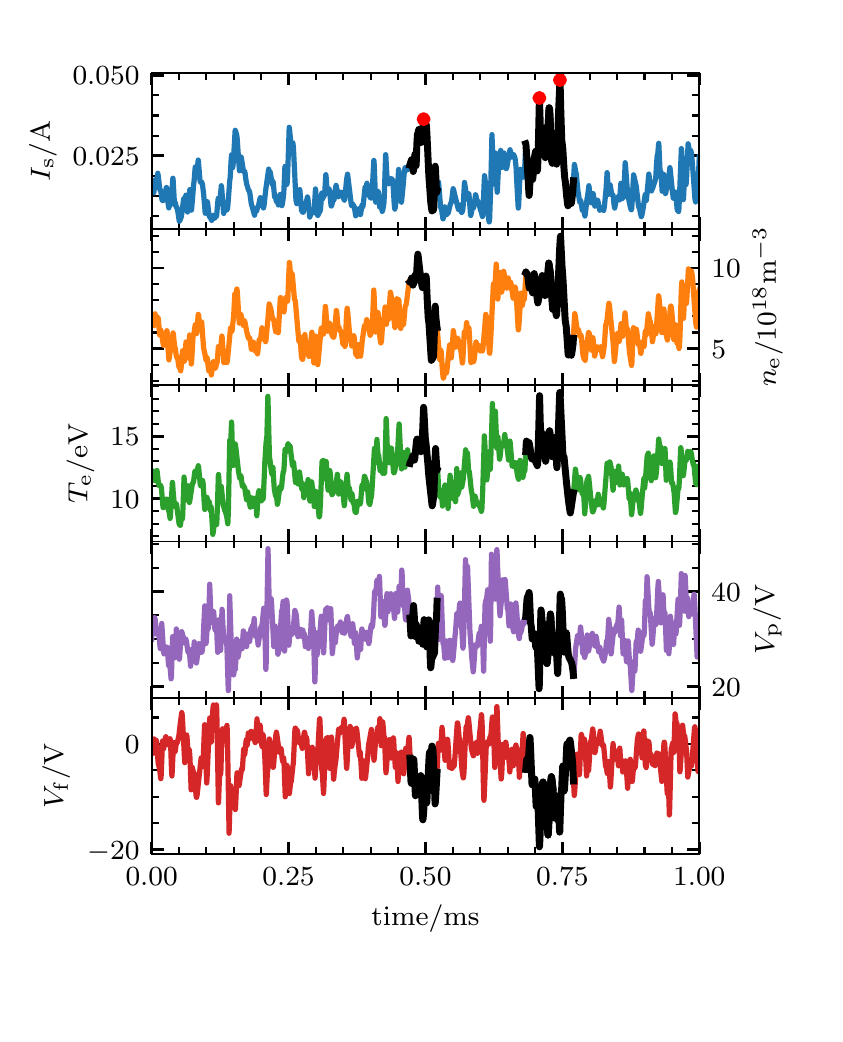}
    \caption{Time series of the ion saturation current, electron density and temperature,
             and plasma and the floating potentials. Local maxima exceeding 2.5 times the
             root mean square value of the $\Is$ time series are marked by red dots.
             The black lines mark $50\,\mus$ long sub-records centered around these maxima.}
    \label{fig:timeseries}
\end{figure}

Figure \ref{fig:timeseries} shows one millisecond long sub-records of the $\Is$, $\nee$,
$\Te$, $\Vp$ and $\Vf$ data time series. Local maxima of the ion saturation current time
series exceeding $2.5$ times the sample root mean square value are marked with a red
circle and $50\, \mus$ long sub-records surrounding these local maxima are marked in black
in all data time series. A visual inspection suggest that large amplitude fluctuations in
the ion saturation current are correlated with similar large amplitude fluctuations in the
electron density and temperature time series. These large-amplitude bursts appear to occur
on a similar time scale. The Pearson sample correlation coefficient for the $\Is$ and $\nee$
time series is given by $R_{\Is, \nee} = 0.91$. This substantiates the approach taken in the
analysis of conventional Langmuir probe data time series, namely that fluctuations in $\Is$
are used as a proxy for fluctuations in $\nee$. Furthermore, we find a sample correlation
coefficient for $\Is$ and $\Te$ given by $R_{\Is, \Te} = 0.83$. This suggests that
fluctuation statistics are similar for these two time series. The plasma potential and the
electron temperature present fluctuations on similar time scales. However, there is no
correlation between large amplitude fluctuations apparent between the two time series.
Fluctuations in the floating potential are anti-correlated to fluctuations in the ion
saturation current, with a Pearson sample correlation coefficient given by 
$R_{\Is, \Vf} = -0.33$. 

%%%%%%%%%%%%%%%%%%%%%%%%%%%%%%%%%%%%%%%%%%%%%%%%%%%%%%%%%%%%%%%%%%%%%%%%%%%%%%%%%%%%%%%%%%%%%%%%%%%
% Figure 01
%%%%%%%%%%%%%%%%%%%%%%%%%%%%%%%%%%%%%%%%%%%%%%%%%%%%%%%%%%%%%%%%%%%%%%%%%%%%%%%%%%%%%%%%%%%%%%%%%%%

%\begin{table}
%    \begin{tabular}{c|c|c|c|c}
%                & $\Is / \mathrm{mA}$   & $\Vf / V$ & $\Te / \mathrm{eV}$   & $\nee / 10^{18} \mathrm{m}^{-3}$  \\ \hline
%        Average & $18$                  & $1.5$     & $14$                  & $6.6$ \\              
%        rms     & $8.5$                 & $5.3$     & $2.7$                 & $1.7$
%    \end{tabular}
%    \caption{Lower order statistics of the $\Is$, $\Vf$, $\nee$ and $\Te$ data time series.}
%    \label{tab:sample_stats_single}
%\end{table}

For further analysis of the data time series we rescale them as to have locally vanishing 
mean and unity variance:
\begin{align}
    \widetilde{\Psi} = \frac{\Psi - \ma{\Psi}}{\mrms{\Psi}}. \label{eq:signal_normalization}
\end{align}
The moving average and moving root mean square time series are computed from samples at 
$t_i = i \triangle_t$ as
\begin{align}
    \ma{\Psi}(t_i)   & = \frac{1}{2r + 1} \sum\limits_{k=-r}^{r} \Psi(t_{i + k}), \label{eq:moving_avg}\\
    \mrms{\Psi}(t_i) &= \left[ \frac{1}{2r + 1} \sum\limits_{k=-r}^{r} \left( \Psi(t_{i + k}) - \ma{\Psi(t_i)} \right)^2 \right]^{1/2}. \label{eq:moving_rms}
\end{align}
where $\triangle_t = 0.3\, \mus$ is the sampling time.
Using a filter radius $r = 16384$, which corresponds to approximately $5\, \mathrm{ms}$,
ensures that both the moving average and the moving root mean square time series feature 
little variation. Indeed, the sample averages of all rescaled time series are approximately 
$10^{-3}$ and their standard deviations deviates from unity by a comparable factor.

\begin{figure}
    % created by mkplot_normalization.py
    \includegraphics{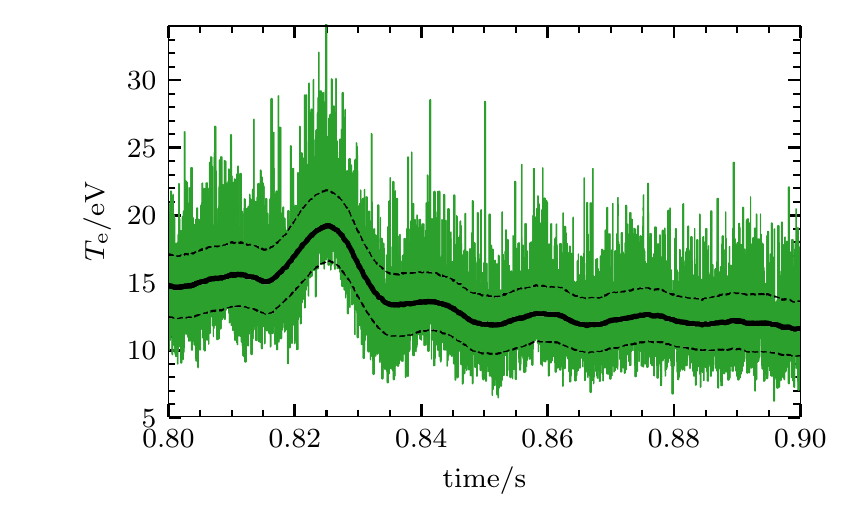}
    \caption{The electron temperature time series (green) and its moving average, defined by
    \Eqnref{moving_avg} (black solid line). The black dashed lines are one moving standard deviation,
    defined by \Eqnref{moving_rms}, above and below the moving average.}
    \label{fig:normalization}
\end{figure}

Figure \ref{fig:normalization} illustrates this rescaling. It shows the $\Te$ time series 
in physical units in green. The moving average, defined by \Eqnref{moving_avg}, is shown
by the solid black line and flanked by the moving root mean square, shown by the dashed
black lines.
While the moving root mean square varies little, between $2.5$ and $3.5\, \mathrm{eV}$,
the moving sample average varies between $12$ and $19\, \mathrm{eV}$. Absorbing these
variations into the normalization of the time series allows to compare samples of the
entire one second long data time series.

All rescaled data time series present non-vanishing sample coefficients of skewness and 
excess kurtosis, or flatness, listed in \Tabref{stats_rescaled}. While the electron 
density and temperature time series feature comparable coefficients of skewness, this 
moment is larger for the ion saturation current. Similarly, the flatness of the ion
saturation current time series is consistently larger than for either electron quantity.
The floating potential features negative coefficients of sample skewness and non-vanishing
coefficients of flatness. On the other hand the plasma potential is skewed towards
positive values and also features positive coefficients of sample kurtosis.

\begin{table}[h!tb]
    % See graphics/stat_single.py
    \begin{tabular}{l|c|c}
        Quantity    & Skewness                          & Flatness \\ \hline
        $\Istilde$  & $1.1 / 1.0 / 1.2 / 1.1$           & $2.0 / 1.7 / 2.5 / 2.1$     \\
        $\Vftilde$  & $-0.23 / -0.23 / -0.83 / -0.64$   & $0.031 / 0.18 / 0.86 / 0.66$ \\
        $\Vptilde$  & $0.53 / 0.64 / 0.76 / 0.60$       & $1.2 / 1.6 / 1.9 / 1.4$ \\
        $\netilde$  & $0.69$                            & $0.79$ \\
        $\Tetilde$  & $0.63$                            & $0.88$ 
    \end{tabular}
    \caption{Sample skewness and flatness of the time series sampled by the
             southwest / northwest / northeast / southeast MLP ($\Istilde$, $\Vftilde$, $\Vptilde$)
             and of the time series averaged over all electrodes ($\netilde$, $\Tetilde$).}
    \label{tab:stats_rescaled}
\end{table}

%%%%%%%%%%%%%%%%%%%%%%%%%%%%%%%%%%%%%%%%%%%%%%%%%%%%%%%%%%%%%%%%%%%%%%%%%%%%%%%%%%%%%%%%%%%%%%%%%%%
% Figure 02
%%%%%%%%%%%%%%%%%%%%%%%%%%%%%%%%%%%%%%%%%%%%%%%%%%%%%%%%%%%%%%%%%%%%%%%%%%%%%%%%%%%%%%%%%%%%%%%%%%%

Compound quantities such as the local electric field and electron particle and heat fluxes are
commonly estimated by combining floating potential and ion saturation current measurements.
An estimator for the radial electric drift velocity is given by 
\begin{align}
    \Vrad = \frac{V^{\mathrm{S}} - V^{\mathrm{N}}}{B \triangle_Z}. \label{eq:vrad_estimator}
\end{align}
Here $B = 4.1\, \mathrm{T}$ gives the magnetic field at the probe head position, and 
$(V^\mathrm{S} - V^\mathrm{N}) / \triangle_Z$ denotes an estimator for the poloidal electric field.
The north and south electrodes of the used probe head are vertically separated by 
$\triangle_Z = 2.2 \times 10^{-3}\, \mathrm{m}$. In the following we estimate the potential at either
poloidal position as the average plasma potential as 
$V^\mathrm{N/S} = (\Vp^{\mathrm{NE/SE}} + \Vp^{\mathrm{NW/SW}} ) / 2$. Using the plasma potential
instead of the floating potential to estimate the electric field includes effects of short-wavelength
electron temperature perturbations on the radial velocity.
Since toroidal plasma drifts may bias electrodes on the same magnetic flux surface to different
electric potentials, the sample mean is subtracted from potential time series used in velocity estimators.
Postulating that there is no stationary convection in the scrape-off layer we further subtract the
moving average from radial velocity time series such that $\mean{\widetilde{\Vrad}}_\mathrm{mv} = 0$.
%RMS(Vrad) = 457.17 m/s, see graphics/stat_combined.py

\begin{table}
    \begin{tabular}{c|c|c|c|c|c}
                %& $\Is / \mathrm{mA}$   & $\Vf / \mathrm{V}$    & $\Vp / \mathrm{V}$    & $\Te / \mathrm{eV}$   & $\nee / 10^{18} \mathrm{m}^{-3}$  \\ \hline
                & $\Is / \mathrm{mA}$   &  $\nee / 10^{18} \mathrm{m}^{-3}$ & $\Te / \mathrm{eV}$   &  $\Vf / \mathrm{V}$   & $\Vp / \mathrm{V}$ \\ \hline
        Average & $18$                  & $6.6$                             & $14$                  & $1.5$                 & $38$                 \\
        rms     & $8.7$                 & $1.8$                             & $2.7$                 & $5.5$                 & $8.2$
        %Average & $18$                  & $1.5$                 & $38$                  & $14$                  & $6.6$ \\              
        %rms     & $8.7$                 & $5.5$                 & $8.2$                 & $2.7$                 & $1.8$
    \end{tabular}
    \caption{Lowest order statistical moments of the $\Is$, $\Vf$, $\Vp$, $\Te$ and $\nee$ data time series.}
    \label{tab:sample_stats_single}
\end{table}

With fast sampling of the electron density and temperature at hand, the radial electron particle
and heat fluxes are estimated as
\begin{align}
    \Gammanhat & = \netilde \Utilde, \label{eq:gamman_estimator}\\ 
    \Gammathat & = \frac{\ma{\nee}}{\mrms{\nee}} \Tetilde  \Utilde
                    + \netilde \frac{\ma{\Te}}{\mrms{\Te}} \Utilde 
                    + \netilde \Tetilde \Utilde
                    .\label{eq:gammat_estimator}
\end{align}
Here, $\netilde$, $\Tetilde$, as well as moving average and moving root mean square time
series denote quantities averaged over all four MLPs. This is done as to use all available
data of the electron temperature as well as to average out outliers. Table 
\ref{tab:sample_stats_single} may be used to convert the amplitude of the estimator time
series to physical units. We note that \Eqsref{gamman_estimator} and
\ref{eq:gammat_estimator} define fluctuation driven fluxes. The total fluctuation driven
heat flux as defined above comprises a conductive contribution, a convective contribution, 
and a contribution driven by triple correlations.

\section{Fluctuation statistics}
\label{sec:single}
%%%%%%%%%%%%%%%%%%%%%%%%%%%%%%%%%%%%%%%%%%%%%%%%%%%%%%%%%%%%%%%%%%%%%%%%%%%%%%%%%%%%%%%%%%%%%%%%%%%
% Figure 03
% created by mkplot_histogram_fit_is_ne_te.py
\begin{figure}
    \includegraphics[width=\textwidth]{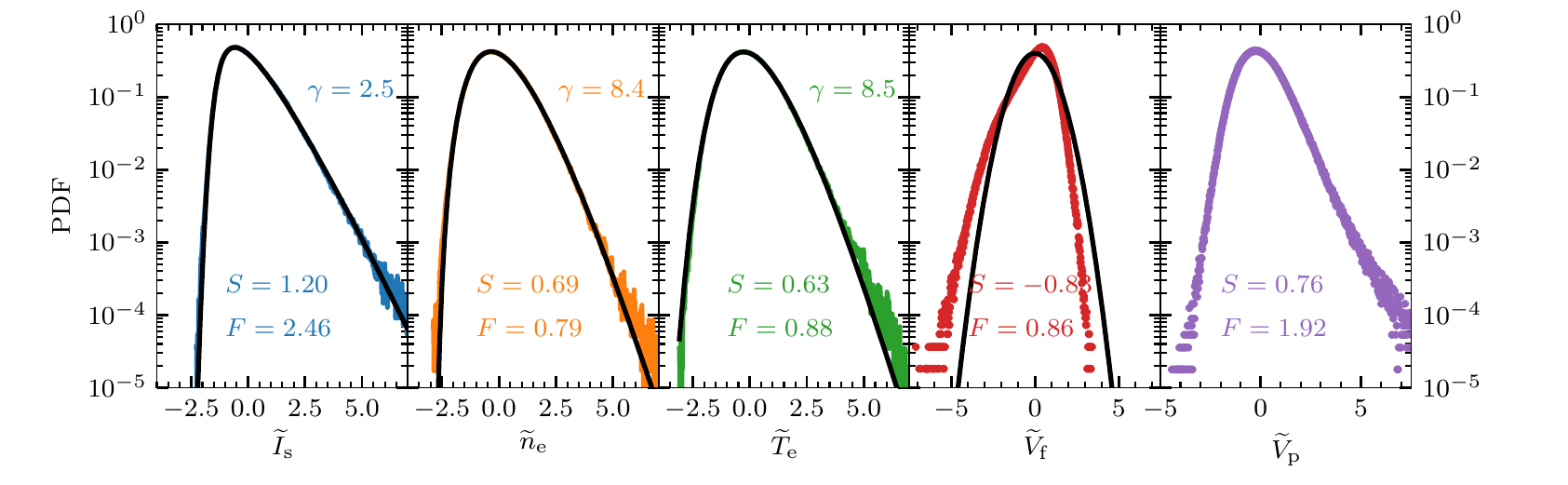}
    \caption{Probability distribution function of the rescaled ion saturation current,
             the electron density and temperature. Compared are least squares fits of 
             the convolution of a Gamma and a normal distribution (black lines) 
             to the PDFs. The two rightmost panels show the PDFs of the floating potential 
             (red dots) compared to a normal distribution (black line) and of the
             plasma potential (purple dots).}
    \label{fig:histogram_is_ne_te}
\end{figure}
%%%%%%%%%%%%%%%%%%%%%%%%%%%%%%%%%%%%%%%%%%%%%%%%%%%%%%%%%%%%%%%%%%%%%%%%%%%%%%%%%%%%%%%%%%%%%%%%%%%

Figure \ref{fig:histogram_is_ne_te} shows PDFs of the rescaled ion saturation current, the electron
density and the electron temperature time series. The two rightmost panels show the PDFs of the floating
and the plasma potential. Least squares fits of the convolution of a normal and a Gamma distribution
to the PDFs of $\Istilde$, $\netilde$, and $\Tetilde$ are shown by black lines. This distribution
arises when assuming that the data time series are due to super-position of uncorrelated exponential
pulses with an exponential amplitude distribution and additive white noise, see appendix A in
\Ref{theodorsen-2016-php}. The shape parameter of this distribution is given by $\gamma = \taud / \tauw$,
where $\taud$ is the pulse duration time and $\tauw$ is the average pulse waiting time. Large values
of $\gamma$ describe time series that are characterized by significant pulse overlap. Realizations
of the process described by \Eqnref{shotnoise} with small values of $\gamma$ feature more isolated
pulses. The signal to noise ratio of the additional white noise is given by $1/\epsilon$. For small
values of $\epsilon$ the signal amplitude is governed by the arrival of exponential pulses, with
additive noise contributing little to the signal amplitude.

% Results from mkplot_histogram_fit_is_ne_te.py
%Is: res_cf =  [ 2.67937065  0.03162134]  error =  [ 0.00790973  0.00068818]
The PDF of the ion saturation current time series features an elevated tail for large amplitude values.
Sample coefficients of skewness and flatness are given by $S = 1.2$ and $F = 2.5$. A least-squares
fit of the prediction by the stochastic model to the PDF yields $\gamma = 2.5$ and
$\epsilon = 3.7 \times 10^{-2}$. This fit describes the PDF well over four decades in normalized
probability.
%
% ne: res_cf =  [  8.61842091e+00   1.00000000e-04]
% Te: res_cf =  [ 11.82364393   0.04785315]  error =  [ 0.23711502  0.00661865]
The PDFs of $\netilde$ and $\Tetilde$ feature a similar shape but with less elevated tails for large,
positive sample values. This is reflected in values of sample skewness and flatness given by $S=0.69$
and $F=0.79$ for $\netilde$ and by $S = 0.63$ and $F=0.88$ for $\Tetilde$. Fitting the prediction by
the stochastic model to the PDF of the sampled data yields $\gamma = 8.4$ and $\epsilon = 0$ for
$\netilde$ and $\gamma = 8.5$ and $\epsilon = 0.13$ for $\Tetilde$. Again, these parameters suggest
a process with significant pulse overlap and little white noise.

Continuing with the PDF of the floating potential we find that negative sample values are more
probable then positive sample values. This is reflected by negative value of the sample skewness,
$S=-0.83$. The PDF deviates from a normal distribution, shown by the black line in the rightmost
panel of \Figref{histogram_is_ne_te}, and reflected by a non-vanishing sample flatness $F=0.86$.
The PDF of the plasma potential features an elevated tail, similar to the PDF of $\Tetilde$. On the
other hand negative $\Vptilde$ samples are more probable than negative $\Tetilde$. Sample skewness
and excess kurtosis are both non-vanishing for the plasma potential.

PDFs of $\Istilde$ and $\Vftilde$ recorded by the other MLPs are qualitatively similar to those shown
here. Interpreting the PDFs with the relationship given by \Eqnref{Isat} one may speculate that the
elevated tail in the ion saturation current PDF is due to simultaneous large amplitude fluctuations
of the electron density and temperature. This issue will be discussed further in the following sections.

\begin{figure}
    \includegraphics{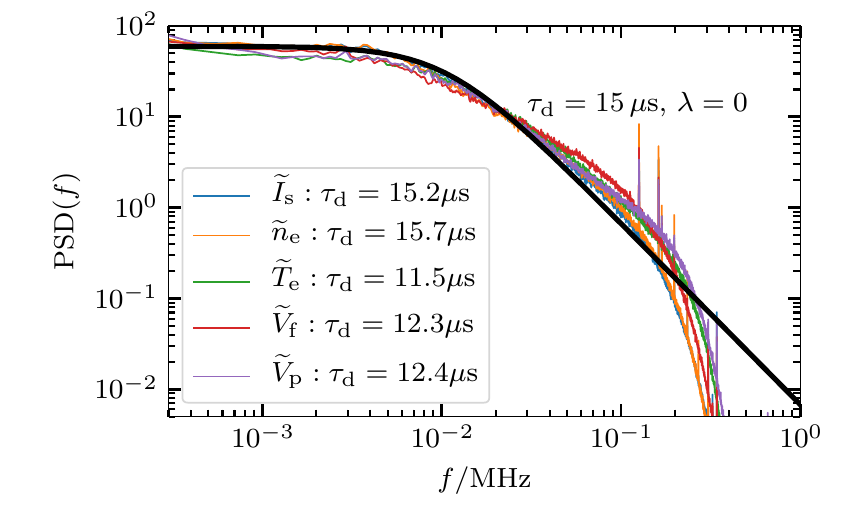}
    \caption{Power spectral density of the ion saturation current, the floating and the
             plasma potential, and the electron density and temperature. The black
             line denotes \Eqnref{shotnoise_psd} with $\taud = 15\, \mus$.}
    \label{fig:psd}
\end{figure}

%%%%%%%%%%%%%%%%%%%%%%%%% Power spectra %%%%%%%%%%%%%%%%%%%%%%%%%%%%%%%%%%%%%%%%%%%%%%%%%%
Figure \ref{fig:psd} shows the power spectral densities (PSDs) of the $\Istilde$, $\netilde$, 
$\Vptilde$, $\Vftilde$ and $\Tetilde$ data time series. They all feature a similar shape, suggesting
that fluctuations in the data time series are due to structures with similar characteristic time scales.
For $f \lesssim 3 \times 10^{-3}\, \mathrm{MHz}$ the PSDs are flat before they roll over to approximately
follow a power law, $f^{-2}$, for $3 \times 10^{-2}\, \mathrm{MHz} \lesssim f \lesssim 0.1\, \mathrm{MHz}$.
For higher frequencies, the PSDs decay even more steep.
A least squares fit of \Eqnref{shotnoise_psd} to the data gives $\taud \approx 15 \mus$
and $\lambda \approx 0$ for all data time series. The black line gives the curve describe by 
\Eqnref{shotnoise_psd} with just this pulse duration time and vanishing pulse rise time. 
Equation (\ref{eq:shotnoise_psd}) states that the flat part of the PSD as well as 
the roll-over frequency is determined by the pulse duration time $\taud$. The pulse 
asymmetry parameter $\lambda$ determines the slope of the PSD after the roll-over. We find 
that the prediction of the stochastic model with parameters found from least squares
fits describe the experimental data well over approximately two decades.

%
%The amplitude of the oscillations for $f \gtrsim 0.1 \mathrm{MHz}$ are less than three
%orders of magnitude lower than the low-frequency part of the PSD. To investigate their 
%origin we generated realizations of the process described by \Eqnref{shotnoise}. A 12 
%point running average filter was applied to this data, similar to the one used to remove
%short timescale fluctuations in the MLP fit input data. While the PSD of the unfiltered
%time series agrees perfectly with the prediction \Eqnref{shotnoise_psd} the
%PSD of the filtered time series shows similar oscillations for high frequencies as those
%seen in \Figref{psd}. We thus conclude that these oscillations are an artifact of the
%running average filter applied to the current samples, used as the input of the MLP fit.

%%%%%%%%%%%%%%%%%%%%%%%%%%%%%%%%%%%%%%%%%%%%%%%%%%%%%%%%%%%%%%%%%%%%%%%%%%%%%%%%%%%%%%%%%%%%%%%%%%%
% Created by mkplot_pspectrum.py
% Figure 04

\begin{figure}
    \includegraphics{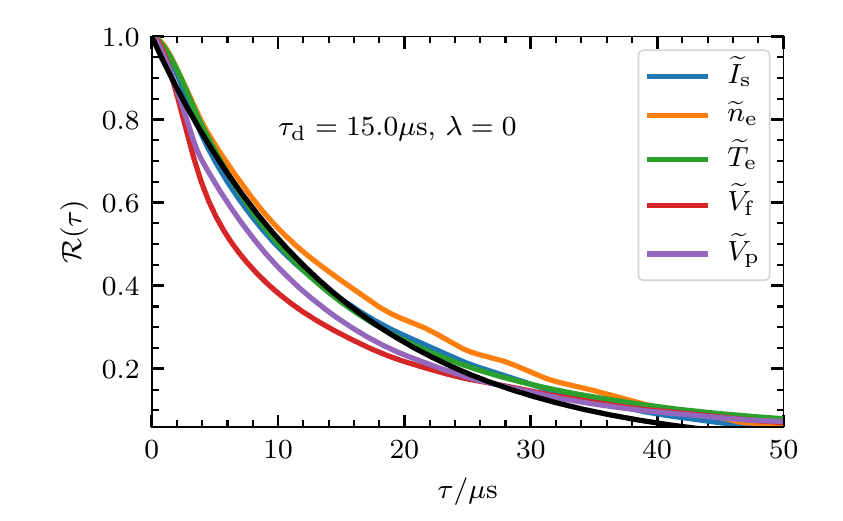}
    \caption{Auto-correlation of the rescaled ion saturation current, electron density
             and temperature as well as the plasma and the floating potential.
             Compared is \Eqnref{shotnoise_acorr} with $\taud = 15\, \mu \mathrm{s}$
             and $\lambda = 0$, denoted by the black line.}
    \label{fig:autocorrelation}
\end{figure}

%%%%%%%%%%%%%%%%%%%%%%%%%% Autocorrelation function %%%%%%%%%%%%%%%%%%%%%%%%%%%%%%%%%%%%%%%%%%%%%%%
Figure \ref{fig:autocorrelation} shows the auto-correlation function for the data time series. The
auto-correlation function of $\Istilde$, $\netilde$, and $\Tetilde$ decay approximately exponentially
for $\tau \lesssim 20\, \mus$. The auto-correlation function of the $\Vftilde$ and $\Vptilde$ data time
series decay faster than exponential. A least-squares fit of \Eqnref{shotnoise_acorr} to the data for
$\tau < 25\, \mus$ gives $\taud \approx 15 \,\mus$ and $\lambda \approx 0$ for $\Istilde$ and $\Tetilde$.
For $\netilde$ a fit yields $\taud \approx 16\, \mus$ and a vanishing pulse asymmetry parameter. 
These parameters agree with the parameters estimated from fits to the PSDs of the data time series.
%Is (array([1.43072775e+01, 1.00000000e-05]), array([[ 1.42128815e-02, -2.20288127e-04],
%       [-2.20288127e-04,  5.50023810e-05]]))
%ne (array([1.63129449e+01, 1.00000000e-05]), array([[ 1.09671825e-02, -1.77141596e-04],
%       [-1.77141596e-04,  2.86718129e-05]]))
%Te (array([1.42865918e+01, 7.89126190e-03]), array([[ 5.11324488e-03, -8.39326922e-05],
%       [-8.39326922e-05,  1.98927828e-05]]))
%Vf (array([1.04952991e+01, 1.00000000e-05]), array([[ 0.04214376, -0.00045409],
%       [-0.00045409,  0.00037023]]))
%Vp (array([1.19190656e+01, 1.00000000e-05]), array([[ 0.02015885, -0.00026476],
%       [-0.00026476,  0.00012859]]))

%%%%%%%%%%%%%%%%%%%%%%%%%%%%%%%%%%%%%%%%%%%%%%%%%%%%%%%%%%%%%%%%%%%%%%%%%%%%%%%%%%%%%%%%%%%%%%%%%%%
% created by mkplot_autocorr_is_ne_te.py
% Figure 05
%%%%%%%%%%%%%%%%%%%%%%%%%%%%%%%%%%%%%%%%%%%%%%%%%%%%%%%%%%%%%%%%%%%%%%%%%%%%%%%%%%%%%%%%%%%%%%%%%%%

\begin{figure}
    \includegraphics{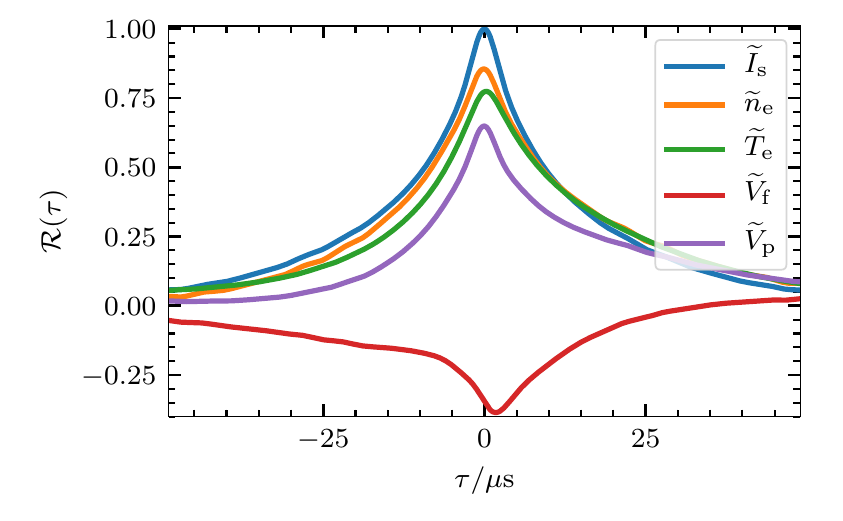}
        \caption{Cross-correlation between all data time series and the ion
                 saturation current.}
    \label{fig:cross_correlation_singleq}
\end{figure}

Figure \ref{fig:cross_correlation_singleq} shows cross-correlation functions between the ion saturation
current and the other data time series. The correlation functions $\mathcal{R}_{\Istilde, \netilde}(\tau)$
and $\mathcal{R}_{\Istilde, \Tetilde}(\tau)$ appear similar in shape. They feature maximum correlation
amplitudes of approximately $0.75$ at nearly vanishing time lag and are slightly asymmetric, with the
correlation amplitude decaying slower for positive than for negative time lags. The cross-correlation
function for the plasma potential, $\mathcal{R}_{\Istilde, \Vptilde}(\tau)$, features a maximal
correlation amplitude of approximately $0.6$ at vanishing time lag. It decays slower to zero for positive
time lags than for negative time lags. The cross-correlation function for the floating potential,
$\mathcal{R}_{\Istilde, \Vftilde}(\tau)$ presents a minimal correlation amplitude of approximately
$-0.4$ at $\tau \approx 2 \mus$. It appears symmetric around the minimum for 
$-5 \mus \lesssim \tau \lesssim 8 \mus$ but decays faster to zero for $\tau > 0$ than for $\tau < 0$
for large lags. Observing that all auto-correlation functions vanish for time lags greater than 
$50\, \mus$ we note that we do not observe any long-range correlations.

%%%%%%%%%%%%%%%%%%%%%%%%%%%%%%%%%%%%%%%%%%%%%%%%%%%%%%%%%%%%%%%%%%%%%%%%%%%%%%%%%%%%%%%%%%%%%%%%%%%
% Figure 06.pdf
% created by mkplot_crosscorr_singleq.py
%%%%%%%%%%%%%%%%%%%%%%%%%%%%%%%%%%%%%%%%%%%%%%%%%%%%%%%%%%%%%%%%%%%%%%%%%%%%%%%%%%%%%%%%%%%%%%%%%%%

Complementary to the auto-correlation function we proceed by studying the time series using the conditional
averaging method \Ref{pecseli-1989}. The conditionally averaged waveform of a signal $\Phi$ is computed
by averaging sub-records centered around local maxima of a reference signal $\Psi$ which exceed a threshold
value, typically taken to be $2.5$ times the time series root mean square value:
\begin{align}
    \mathcal{C}_{\widetilde{\Phi}, \widetilde{\Psi} }(\tau) & = 
        \langle \widetilde{\Phi}(\tau) | \widetilde{\Psi}(\tau = 0) > 2.5 ,\,  \widetilde{\Psi}'(0) = 0 \rangle. \label{eq:condavg}
\end{align}
Here the prime denotes a derivative. To ensure that the conditionally averaged waveform is computed
from independent samples, the local maxima are required to be separated by the same interval length
on which \Eqnref{condavg} is computed. For the data sets at hand we choose
$-25\, \mus \leq \tau \leq 25 \, \mus$.

\begin{figure}
    \includegraphics{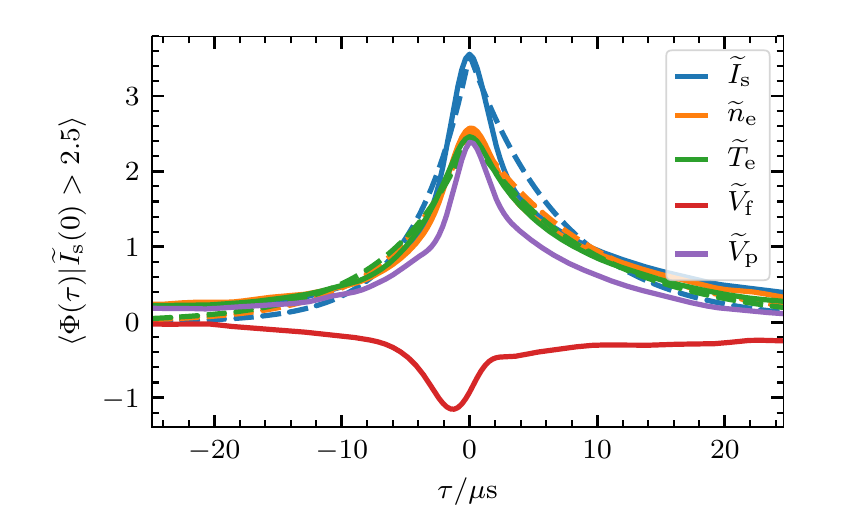}
    \caption{Conditionally averaged waveforms of the data time series, centered 
             around large amplitude maxima in the ion saturation current time 
             series. Dashed lines show least squares fits of a two-sided 
             exponential waveform, given by \Eqnref{pulseshape}, to the 
             conditionally averaged waveforms.}
    \label{fig:condavg_is}
\end{figure}

Figure \ref{fig:condavg_is} shows the conditionally averaged waveform of the
$\Istilde$, $\netilde$, $\Tetilde$, $\Vftilde$ and $\Vptilde$ data time series, using $\Istilde$ as
a reference signal. Approximately $4000$ maxima are detected in the $\Istilde$ time series. The
conditionally averaged waveform of the ion saturation current is strongly peaked and decays faster
then exponentially to zero for large time lags. The average amplitude of the local ion saturation
current maxima is approximately three times the time series root mean square value. The conditionally
averaged waveforms of the electron density and temperature are both well approximated by a two-sided
exponential function. The maxima of their waveforms are approximately two times the root mean square
value of their respective time series. The conditionally averaged waveform of the $\Vptilde$ time
series appears triangular, with the maxima in phase with local maxima of the $\Istilde$ time series.
The conditionally averaged waveform of the floating potential presents a negative peak with an
amplitude of approximately $-1$, occurring at  $\tau \approx -2\, \mus$.
Compared to the averaged waveforms are least square fits of a two-sided exponential waveform, given
by \Eqnref{pulseshape}, to the data, marked by dashed lines in \Figref{condavg_is}. Table 
\ref{tab:taurf_fits} lists their fit parameters. The average waveform duration time is between $12$
and $16\, \mus$, comparable to $\taud$ estimated by fits to the auto-correlation function and power
spectral densities of the signals. The pulse asymmetry parameter for all fits is given by approximately
$0.4$.

%%%%%%%%%%%%%%%%%%%%%%%%%%%%%%%%%%%%%%%%%%%%%%%%%%%%%%%%%%%%%%%%%%%%%%%%%%%%%%%%%%%%%%%%%%%%%%%%%%%
% Figure 07
% Created by mkplot_condavg_is.py
%%%%%%%%%%%%%%%%%%%%%%%%%%%%%%%%%%%%%%%%%%%%%%%%%%%%%%%%%%%%%%%%%%%%%%%%%%%%%%%%%%%%%%%%%%%%%%%%%%%

\begin{table}[h!tb]
    \begin{tabular}{c|c|c|c}
        Waveform                    & $\langle \Istilde | \Istilde(0) > 2.5 \rangle$ & $\langle \netilde | \Istilde(0) > 2.5 \rangle$   & $\langle \Tetilde | \Istilde(0) > 2.5 \rangle$    \\ \hline
        $\taud / \mus$              & $11$                                           & $16$                                             & $16$                                              \\
        $\lambda = \taur / \taud$   & $0.37$                                         & $0.37$                                           & $0.40$                                            \\

    \end{tabular}
    \caption{Duration time of the last squares fits shown in \Figref{condavg_is}
             and the waveform asymmetry parameter $\lambda$.}
    \label{tab:taurf_fits}
\end{table}

%%%%%%%%%%%%%%%%%%%%%%%%%%%%%%%%%%%%%%%%%%%%%%%%%%%%%%%%%%%%%%%%%%%%%%%%%%%%%%%%%%%%%%%%%%%%%%%%%%%
% Created by mkplot_tauwait.py
% Figure 08
\begin{figure}
    \includegraphics{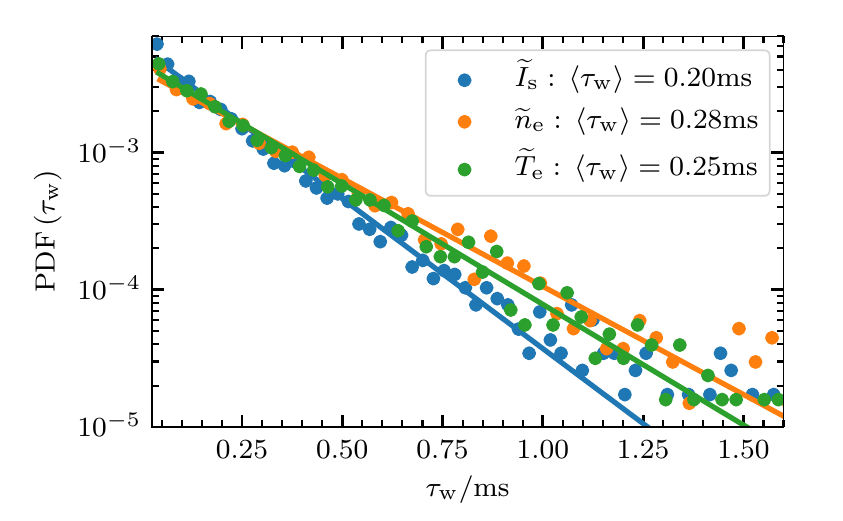}
    \caption{PDFs of the waiting times between large-amplitude bursts in the 
         $\nee$, $\Te$, and $\Is$ time series, recorded by the southwest MLP.
         The full lines show PDFs of a truncated exponential distribution
         with a scale parameter given by a maximum likelihood estimate to
         the data points.}
    \label{fig:tauwait_ne_te_is}
\end{figure}
%%%%%%%%%%%%%%%%%%%%%%%%%%%%%%%%%%%%%%%%%%%%%%%%%%%%%%%%%%%%%%%%%%%%%%%%%%%%%%%%%%%%%%%%%%%%%%%%%%%

Computing the time lag between successive, conditional maxima of the time series yields the waiting
time statistics for large-amplitude events.  
% ne: 2624 bursts
% te: 2915 bursts
% is: 3347 bursts
% scale: 366.915512
% scale: 332.829976
% scale: 246.088943
Figure \ref{fig:tauwait_ne_te_is} shows the waiting time PDF of the $\Istilde$, $\netilde$ and 
$\Tetilde$ time series. Compared are PDFs of exponentially distributed variables. Their scale
parameter is given by a maximum likelihood estimate of the waiting time distribution. The resulting
distributions describe the data well over approximately two decades in probability. Average waiting
times are given by approximately $0.25\, \mathrm{ms}$ for $\Tetilde$ and $0.28\, \mathrm{ms}$ for
$\netilde$. The average waiting time between large-amplitude bursts in $\Istilde$ is approximately
$0.20\, \mathrm{ms}$. We note that the exact numerical values depend slightly on the threshold value
and the conditional window length used.

\begin{figure}
    \includegraphics{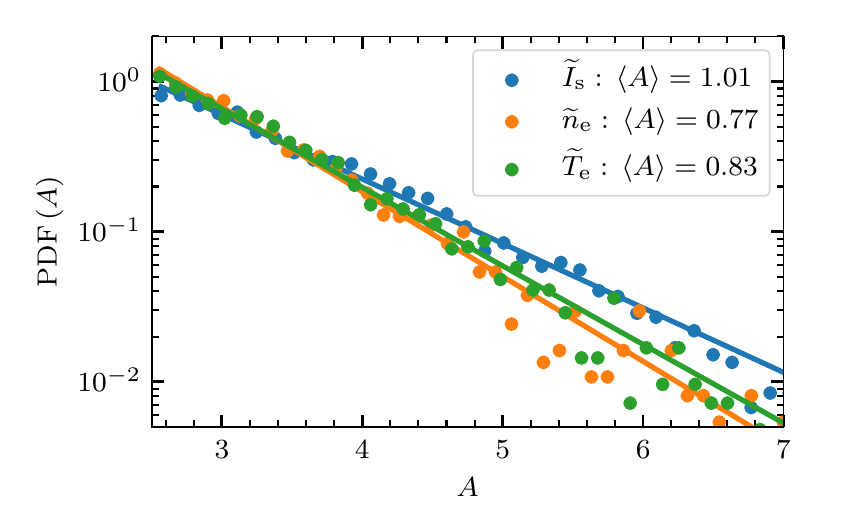}
    \caption{PDF of large amplitude local maxima in the rescaled time series. Full lines 
             show the PDFs of a truncated exponential distribution with a scale parameter 
             given by a maximum likelihood estimate to the data time series.}
    \label{fig:burstamps_ne_te_is}
\end{figure}

The PDFs of the signals local maxima are shown in \Figref{burstamps_ne_te_is}. As for the average
waiting times, the PDFs are well described by a truncated exponential distribution. The scale parameter,
found by maximum likelihood estimates of the data, are given by $\mean{A} \approx 1.0$ for $\Istilde$,
$\mean{A} \approx 0.77$ for $\netilde$, and by $\mean{A} \approx 0.83$ for $\Tetilde$.
Given the threshold amplitude of $2.5$, this translates to an average burst amplitude of the rescaled
signals between $3.3$ and $3.5$ times the root-mean-square value of the data time series, consistent
with the amplitude of the conditionally averaged waveforms shown in \Figref{burstamps_ne_te_is}.

%%%%%%%%%%%%%%%%%%%%%%%%%%%%%%%%%%%%%%%%%%%%%%%%%%%%%%%%%%%%%%%%%%%%%%%%%%%%%%%%%%%%%%%%%%%%%%%%%%%
% Created by mkplot_burstamp.py
% Figure 09
%%%%%%%%%%%%%%%%%%%%%%%%%%%%%%%%%%%%%%%%%%%%%%%%%%%%%%%%%%%%%%%%%%%%%%%%%%%%%%%%%%%%%%%%%%%%%%%%%%%

\section{Radial velocity and fluxes}
\label{sec:multi}

%%%%%%%%%%%%%%%%%%%%%%%%%%%%%%%%%%%%%%%%%%%%%%%%%%%%%%%%%%%%%%%%%%%%%%%%%%%%%%%%%%%%%%%%%%%%%%%%%%%
% Created by mkplot_timetraces_multi.py
% Figure 10
\begin{figure}
    \includegraphics{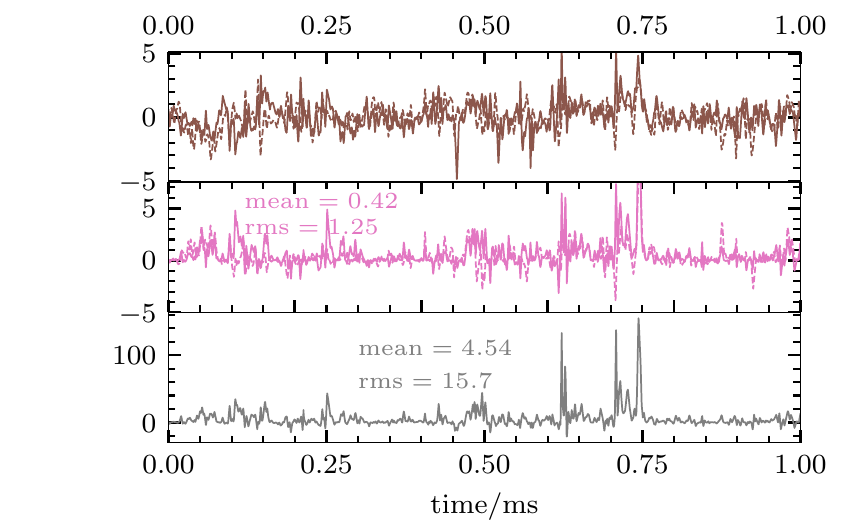}
    \caption{Estimators for the radial velocity (upper panel) and the radial electron particle, and heat 
             flux (middle and lower panel). The full (dashed) line in the upper panel denotes the
             radial velocity estimated from $\Vp$ ($\Vf$) samples. The full (dashed) line in
             the middle panel denotes the radial electron flux esimated from $\netilde$ and
             $\Utilde_{V_\mathrm{p}}$ ($\Istilde$ and $\Utilde_{V_\mathrm{f}}$) samples.
             The time interval is identical to the one 
             presented in \Figref{timeseries}.}
    \label{fig:timetraces_multi}
\end{figure}
%%%%%%%%%%%%%%%%%%%%%%%%%%%%%%%%%%%%%%%%%%%%%%%%%%%%%%%%%%%%%%%%%%%%%%%%%%%%%%%%%%%%%%%%%%%%%%%%%%%

In the following the statistical properties of the radial velocity and electron particle and heat
fluxes are discussed. Figure \ref{fig:timetraces_multi} shows $1\, \mathrm{ms}$ long time series
of the estimators given by \Eqsref{vrad_estimator} - (\ref{eq:gammat_estimator}), computed on the
same time interval as the time series shown in \Figref{timeseries}. The full (dashed) line in the
upper panel denotes the radial velocity estimated from $\Vp$ ($\Vf$) samples. In the middle panel
the full (dashed) line denotes the radial electron flux estimated from $\nee$ and $\Vp$ 
($\Is$ and $\Vf$) samples.
The radial velocity time series show fluctuations on a similar time scale as seen for the time series
shown in \Figref{timeseries}. There is no qualitative difference between $\Utilde$ estimated from the
floating potential and from the plasma potential. Both positive and negative local maxima appear with
nearly equal frequency, not exceeding $5$ in normalized units. The $\Gammanhat$ time series feature
predominantly positive fluctuation amplitudes on a similar temporal scale as the $\Utilde$ time series.
Using ion saturation current and floating potential to estimate the particle flux yields almost
indistinguishable estimator samples. The sample mean and root mean square value are given by $0.42$
($0.48$) and $1.25$ ($1.20$) respectively, using $\nee$ and $\Vp$ ($\Is$ and $\Vf$) samples.
The radial heat flux time series $\Gammathat$ features large-amplitude bursts exceeding $80$ in
normalized units. Large-amplitude temperature fluctuations, which appear in phase with large-amplitude
particle flux events, give rise to this large fluctuation level. Only few large, negative heat flux
events are recorded.
 
%%%%%%%%%%%%%%%%%%%%%%%%%%%%%%%%%%%%%%%%%%%%%%%%%%%%%%%%%%%%%%%%%%%%%%%%%%%%%%%%%%%%%%%%%%%%%%%%%%%
% Created by mkplot_condavg_vrad.py
% Figure 11
\begin{figure}
    \includegraphics{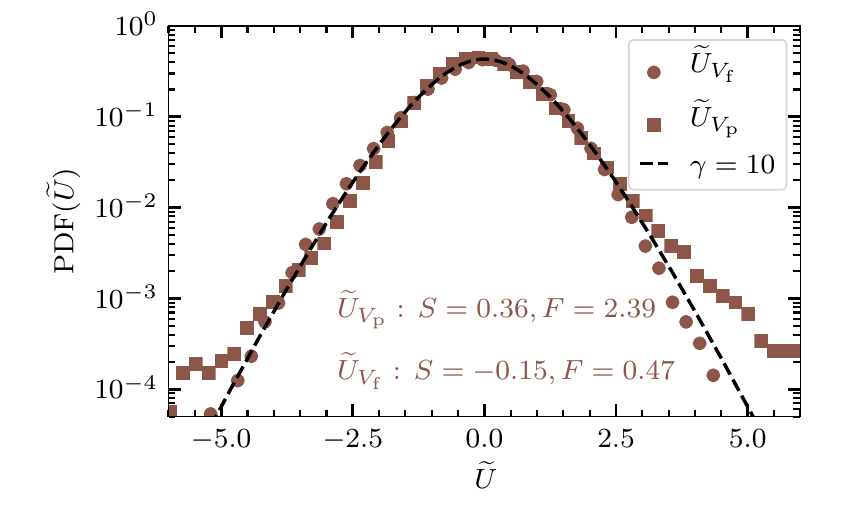}
    \caption{PDF of the rescaled radial electric drift velocity. Compared is the 
             PDF predicted from the stochastic model, given by Eq.(A9) \Ref{theodorsen-2016-ppcf}, 
             for a Laplace distribution of pulse amplitudes.}
    \label{fig:histogram_vrad}
\end{figure}
%%%%%%%%%%%%%%%%%%%%%%%%%%%%%%%%%%%%%%%%%%%%%%%%%%%%%%%%%%%%%%%%%%%%%%%%%%%%%%%%%%%%%%%%%%%%%%%%%%%

Figure \ref{fig:histogram_vrad} presents the PDF of the radial velocity estimator given by
\Eqnref{vrad_estimator}. The PDF of $\Utilde_{\Vftilde}$ appears symmetric with exponential tails for
both positive and negative sample values, compatible with $S=-0.15$ and $F=0.47$. The PDF of
$\Utilde_{\Vptilde}$ is almost identical to the PDF computed using floating potential measurements,
but notably features an elevated tail for large amplitude samples $\Utilde_{\Vptilde} \gtrsim 2.5$.
A correlation analysis of samples 
$\Vf^{\mathrm{S}} - \Vf^{\mathrm{N}}$ and $\Te^{\mathrm{S}} - \Te^{\mathrm{N}}$ showed no correlation
between large-amplitude potential differences to large amplitude electron temperature differences,
which may have explained this artifact in the PDF. The coefficient of sample skewness for 
$\Utilde_{\Vftilde}$ is slightly negative, while the elevated tail of $\mathrm{PDF}(\Utilde_{\Vptilde})$
yields a slightly positive coefficient of sample skewness.
Compared to the PDF is the probability distribution function of the process defined by \Eqnref{shotnoise}
with Laplace distributed pulse amplitudes, which allows for positive as well as negative pulse amplitudes
\Ref{theodorsen-2016-ppcf, theodorsen-2018-ppcf}. Estimating $\gamma$ by a least squares fit to the
PDF of $\Utilde_{\Vftilde}$ yields $\gamma \approx 10$. This value is comparable with the intermittency
parameter for the $\netilde$ and $\Tetilde$ data time series and is larger by a factor of approximately
4 than for the $\Istilde$ time series.

%%%%%%%%%%%%%%%%%%%%%%%%%%%%%%%%%%%%%%%%%%%%%%%%%%%%%%%%%%%%%%%%%%%%%%%%%%%%%%%%%%%%%%%%%%%%%%%%%%%
% Figure 12
\begin{figure}
    \includegraphics{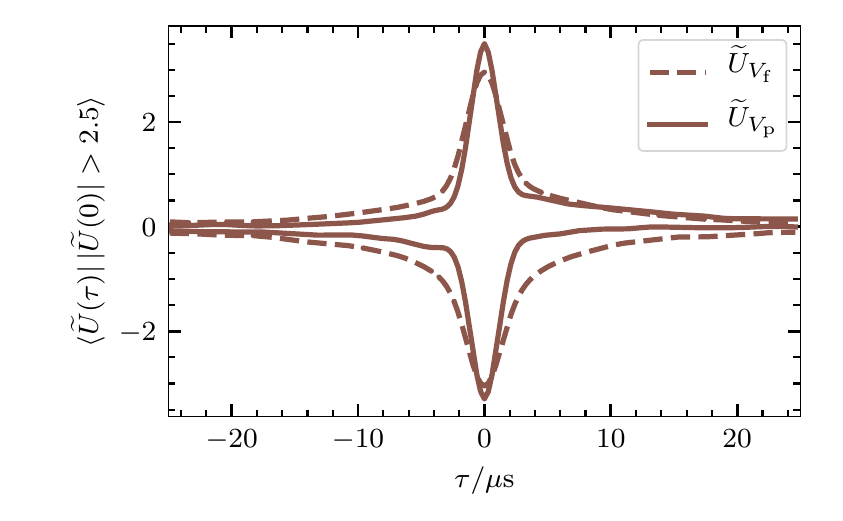}
    \caption{Conditionally averaged waveform of local extrema in the radial
    velocity time series.}
    \label{fig:condavg_vrad}
\end{figure}
%%%%%%%%%%%%%%%%%%%%%%%%%%%%%%%%%%%%%%%%%%%%%%%%%%%%%%%%%%%%%%%%%%%%%%%%%%%%%%%%%%%%%%%%%%%%%%%%%%%

The auto-conditionally averaged waveform of large-amplitude velocity fluctuations, computed from
approximately $3000$ events, are shown in \Figref{condavg_vrad}. The average waveform is approximately
triangular for the $\Utilde_{\Vptilde}$ time series while it is less peaked for the $\Utilde_{\Vftilde}$
time series. The duration time of both waveforms is approximately $5\, \mus$, smaller by a factor of
three than the conditionally averaged waveforms of the electron density and temperature.

\begin{figure}
    \includegraphics{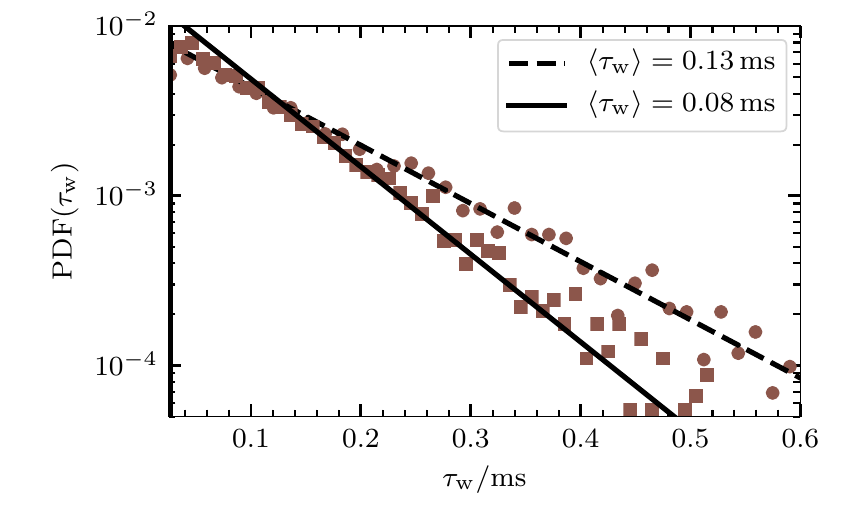}
    \caption{PDF of the waiting times between successive, positive or negative, extrema 
             in the radial velocity time series. Square (circle) plot markers denote
             the radial velocity estimated from $\Vp$ ($\Vf$) samples. Compared are
             exponential distributions with a scale parameter given by 
             a maximum likelihood estimate of the waiting times.}
    \label{fig:tauwait_vrad}
\end{figure}

The PDF of the waiting times between local extrema in the $\Utilde$ time series, both positive and
negative, are shown in \Figref{tauwait_vrad}. Compared are PDFs of exponentially distributed variables
with scale parameters given by
$\mean{\tauw} = 0.08\, \mathrm{ms}$ for $\Utilde_{\Vptilde}$ and by
$\mean{\tauw} = 0.13\, \mathrm{ms}$ for $\Utilde_{\Vftilde}$. These parameters have been estimated
by a maximum likelihood estimate of the respective waiting time data. The resulting distributions
describes the waiting times well over approximately two decades in probability. Varying the minimum
separation between detected local extrema changes the average waiting time only little since positive
and negative maxima are detected independently of each other.

%%%%%%%%%%%%%%%%%%%%%%%%%%%%%%%%%%%%%%%%%%%%%%%%%%%%%%%%%%%%%%%%%%%%%%%%%%%%%%%%%%%%%%%%%%%%%%%%%%%
% Created by mkplot_condavg_vrad.py
% Figure 13
%%%%%%%%%%%%%%%%%%%%%%%%%%%%%%%%%%%%%%%%%%%%%%%%%%%%%%%%%%%%%%%%%%%%%%%%%%%%%%%%%%%%%%%%%%%%%%%%%%%

\begin{figure}
    \includegraphics{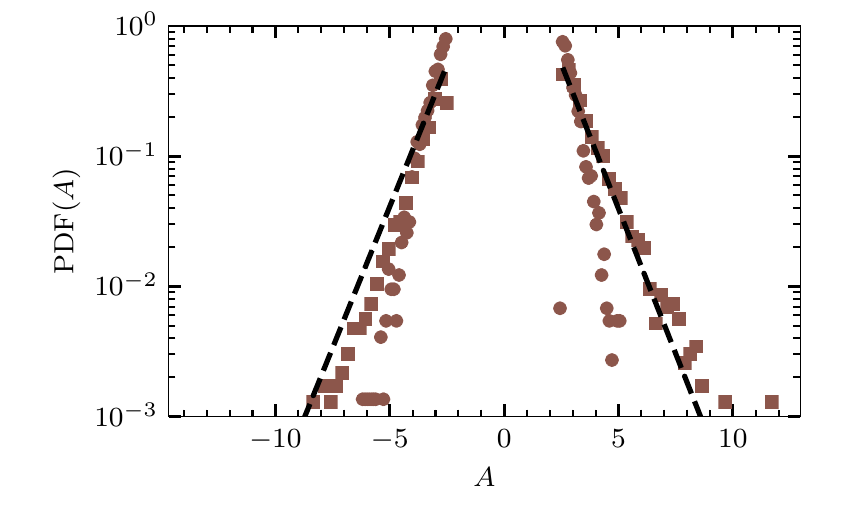}
    \caption{PDF of the local maxima in the radial velocity time series. 
             Square (circle) plot markers denote the radial velocity estimated from
             $\Vp$ ($\Vf$) samples. Compared are
             fits on exponential distribution multiplied by a factor $1/2$, for
             $A > 2.5$ and $A < 2.5$.}
    \label{fig:burstamp_vrad}
\end{figure}

The PDF of the local extrema is shown in \Figref{burstamp_vrad}. Compared to the positive and negative
legs of the distribution are truncated exponential distributions for $|A| > 2.5$. Each distribution
has been multiplied by $1/2$ to normalize the integral of both PDFs to unity. A least squares fit to
the $\Utilde_{\Vptilde}$ data yields a scale parameter of approximately $1$
for both positive and negative amplitudes.
Together with the result that the waiting times between large local maxima of the 
velocity time series are well described by an exponential distribution, these findings
corroborate the hypothesis to interpret the radial velocity time series as a super-position
of uncorrelated pulses described by \Eqnref{shotnoise}.

%%%%%%%%%%%%%%%%%%%%%%%%%%%%%%%%%%%%%%%%%%%%%%%%%%%%%%%%%%%%%%%%%%%%%%%%%%%%%%%%%%%%%%%%%%%%%%%%%%%
% Created by mkplot_condavg_vrad.py
% Figure 14
%%%%%%%%%%%%%%%%%%%%%%%%%%%%%%%%%%%%%%%%%%%%%%%%%%%%%%%%%%%%%%%%%%%%%%%%%%%%%%%%%%%%%%%%%%%%%%%%%%%

\begin{figure}
    \includegraphics{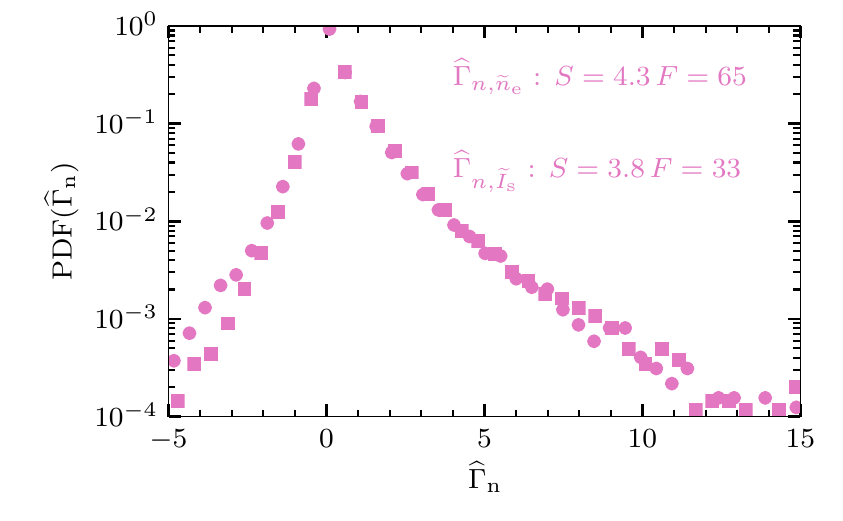}
    \caption{PDF of the normalized radial electron flux. Square (circle) plot markers
             denote the radial electron flux estimated using $\netilde$ and $U_{\Vp}$
             ($\Istilde$ and $U_{\Vf}$) samples.}
    \label{fig:histogram_gamman}
\end{figure}

Figure \ref{fig:histogram_gamman} presents the PDF of the radial particle fluxes computed using either
$\nee$ and $\Vp$ samples or $\Is$ and $\Vf$ samples. The PDFs are almost indistinguishable. They are
strongly peaked at zero and feature non-exponential tails for both positive and negative sample values.
Positive sample values have a much higher probability than negative sample values. This is reflected
by coefficients of sample skewness and excess kurtosis given by $S=4.3$ ($3.8$) and $F=65$ ($33$),
respectively.

%%%%%%%%%%%%%%%%%%%%%%%%%%%%%%%%%%%%%%%%%%%%%%%%%%%%%%%%%%%%%%%%%%%%%%%%%%%%%%%%%%%%%%%%%%%%%%%%%%%
% Created by mkplot_histogram_gamman.py
% Figure 15
%%%%%%%%%%%%%%%%%%%%%%%%%%%%%%%%%%%%%%%%%%%%%%%%%%%%%%%%%%%%%%%%%%%%%%%%%%%%%%%%%%%%%%%%%%%%%%%%%%%

%%%%%%%%%%%%%%%%%%%%%%%%%%%%%%%%%%%%%%%%%%%%%%%%%%%%%%%%%%%%%%%%%%%%%%%%%%%%%%%%%%%%%%%%%%%%%%%%%%%
% Created by mkplot_histogram_gammat.py
% Figure 16
\begin{figure}
    \includegraphics{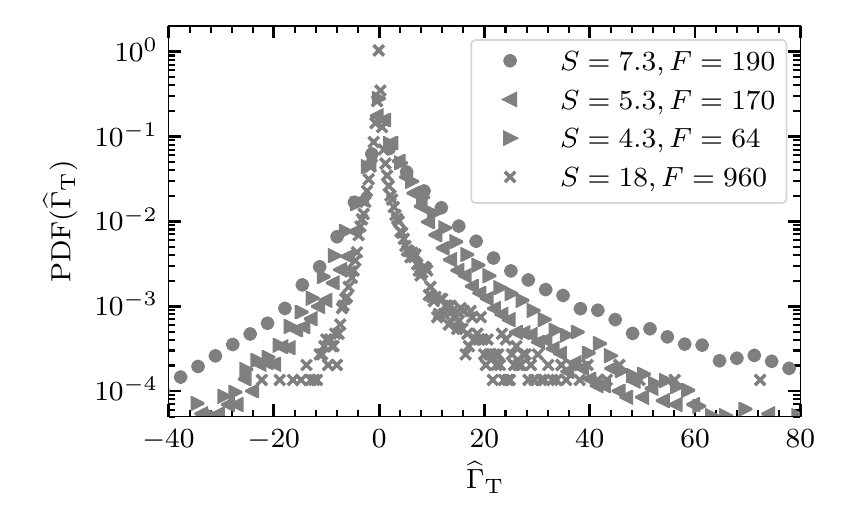}
    \caption{PDF of the total radial heat flux (circle), the conductive (triangle left),
             convective heat fluxes (triangle right), and the heat flux due to triple 
             correlations(cross).}
    \label{fig:histogram_gammat}
\end{figure}
%%%%%%%%%%%%%%%%%%%%%%%%%%%%%%%%%%%%%%%%%%%%%%%%%%%%%%%%%%%%%%%%%%%%%%%%%%%%%%%%%%%%%%%%%%%%%%%%%%%

%
The PDF of the radial heat flux, shown by circles in \Figref{histogram_gammat}, presents a similar
shape with heavy tails for large sample values. Sample coefficients of skewness and flatness are
given by $S = 7.3$ and $F = 190$. Also shown are PDFs of the conductive heat flux
$(\ma{\nee} / \mrms{\nee} )\, \Tetilde\, \Utilde$ (triangle left),
the convective heat flux,
$\netilde (\ma{\Te} / \mrms{\Te} )\, \Utilde$ (triangle right),
and triple correlations 
$\netilde\, \Tetilde\, \Utilde$ (cross).
PDFs of the conductive and convective heat fluxes appear similar in shape as the total heat flux.
However, large-amplitude convective heat flux samples occur more frequently than conductive heat flux
samples of equal magnitude. The PDF of the heat flux due to triple correlations is strongly peaked
for small amplitudes and skewed towards positive sample values. 

\begin{table}[h!tb]
    % See graphics/stat_combined.py
    %tilde(ne) tilde(U_Vp): mean = 2.905527e+20, std = 8.804409e+20
    %tilde(Is) tilde(U_Vf): mean = 1.087924e+20, std = 2.732917e+20
    %Entire Gamma_T: mean = 7.017341e+21, std = 2.490143e+22
    %Conductive: <n> * tilde(T) * tilde(U): mean = 2.674219e+21, rms=9.489370e+21
    %Convective: <T> * tilde(T) * tilde(U): mean = 3.920237e+21, rms=1.217035e+22
    %Triple corr: tilde(n) * tilde (T) * tilde(U): mean = 4.228857e+20, rms=5.527661e+21

    \begin{tabular}{c|c|c|c|c|c}
    & $\Gamman$   
    & $\Gammat$
    & $\ma{\nee}\, \Tetilde\, \Utilde$ 
    & $\ma{\Te}\, \netilde\, \Utilde$ 
    & $\netilde\, \Tetilde\, \Utilde$ \\ \hline
    Average $10^{20}\, \mathrm{m}^{-2} \mathrm{s}^{-1}$            & $2.9$  & $70\, \mathrm{eV}$  & $27\, \mathrm{eV}$ & $39\, \mathrm{eV}$ & $4.2\, \mathrm{eV}$ \\ 
    Root mean square $10^{20}\, \mathrm{m}^{-2} \mathrm{s}^{-1}$   & $8.8$  & $250\, \mathrm{eV}$ & $95\, \mathrm{eV}$ & $120\, \mathrm{eV}$ & $55\, \mathrm{eV}$ \\  
    \end{tabular}
    \caption{Sample average and root mean square value of the contributions to the 
             radial fluxes.}
    \label{tab:flux_contributions}
\end{table}

The sample averages and root mean square values of the various contributions to the total heat flux
are listed in \Tabref{flux_contributions}. This data shows that $38\%$ of the total fluctuation driven
heat flux is due to conduction, $56\%$ due to convection, and $6\%$ due to triple correlations.
For both the particle and the total heat flux we find that their root mean square value is approximately
two-three times their mean value. This also holds for the conductive and the convective heat fluxes.
The relative fluctuation level of the heat flux due to triple correlations is approximately 12.

%%%%%%%%%%%%%%%%%%%%%%%%%%%%%%%%%%%%%%%%%%%%%%%%%%%%%%%%%%%%%%%%%%%%%%%%%%%%%%%%%%%%%%%%%%%%%%%%%%%
% Created by mkplot_jointdf_vrad_ne.py
% Figure 17
\begin{figure} 
    \includegraphics{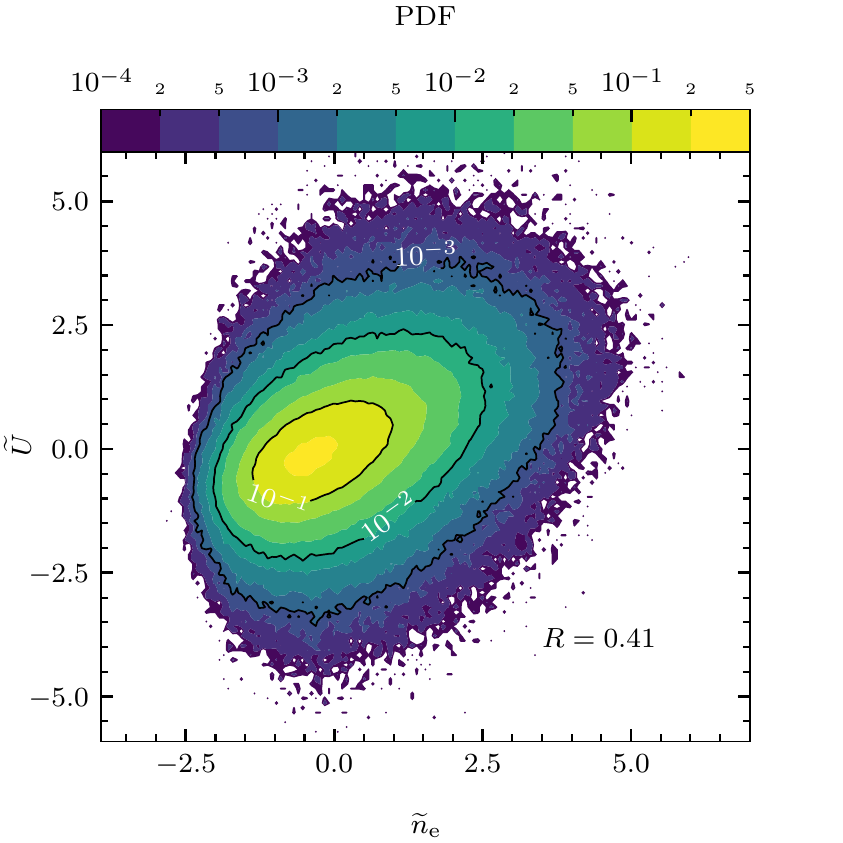}
    \caption{Joint PDF of the radial velocity and the electron density fluctuations.}
    \label{fig:jointpdf_vrad_ne}
\end{figure}
%%%%%%%%%%%%%%%%%%%%%%%%%%%%%%%%%%%%%%%%%%%%%%%%%%%%%%%%%%%%%%%%%%%%%%%%%%%%%%%%%%%%%%%%%%%%%%%%%%%

We continue by discussing the correlations between the electron density and temperature
and the radial velocity fluctuation time series. Figure \ref{fig:jointpdf_vrad_ne} presents
the joint PDF of the fluctuating radial velocity and electron
density. The linear sample correlation coefficient is given by $R=0.41$, consistent with
the slightly tilted shape of the ellipsoids capturing probabilities less than $10^{-1}$,
$10^{-2}$ and $10^{-3}$. Large-amplitude fluctuations are enclosed by equi-probability 
ellipsoids whose semi-minor axis increases with decreasing probability. Negative
large-amplitude velocity fluctuations, $\Utilde \lesssim -2.5$, appear in phase 
with small positive and negative density fluctuations. Positive, large amplitude density 
fluctuations, $\netilde \gtrsim 2.5$, appear on average in phase with positive velocity 
fluctuations. Negative density fluctuations appear on average with vanishing velocity 
fluctuations while positive velocity fluctuations, $\Utilde \gtrsim 2.5$ appear on average 
in phase with positive density fluctuations.

%%%%%%%%%%%%%%%%%%%%%%%%%%%%%%%%%%%%%%%%%%%%%%%%%%%%%%%%%%%%%%%%%%%%%%%%%%%%%%%%%%%%%%%%%%%%%%%%%%%
% Created by mkplot_jointdf_vrad_te.py
% Figure 18
\begin{figure}
    \includegraphics{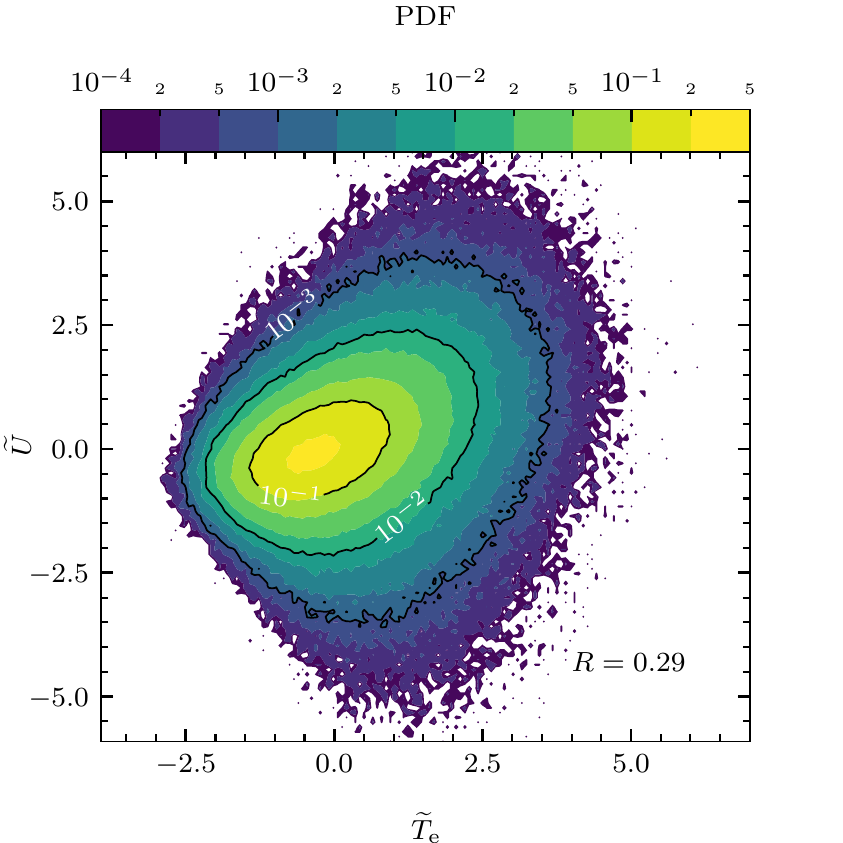}
    \caption{Joint probability distribution function of the radial velocity and electron 
             temperature fluctuations.}
    \label{fig:jointpdf_vrad_te} 
\end{figure}
%%%%%%%%%%%%%%%%%%%%%%%%%%%%%%%%%%%%%%%%%%%%%%%%%%%%%%%%%%%%%%%%%%%%%%%%%%%%%%%%%%%%%%%%%%%%%%%%%%%

The joint PDF of the fluctuating radial velocity and the electron temperature, shown in 
\Figref{jointpdf_vrad_te}, features some qualitative similarities to the joint PDF of the velocity
and density fluctuations. Small-amplitude fluctuations are correlated, captured by a tilted ellipsoid
for a joint probability approximately less than $0.1$. The sample correlation coefficient for the
time series is given by $R=0.29$. Large-amplitude fluctuations are captured by equi-probability
contours whose shape increasingly deviates from an ellipse with decreasing probability. Especially
are large, negative velocity fluctuations observed which are in phase with small, positive temperature
fluctuations. Large negative temperature fluctuations are in phase with small velocity fluctuations,
$\Utilde \approx 0$, with less scatter than observed for the density fluctuations. Large positive
temperature fluctuations are on average in phase with positive temperature fluctuations, also with
larger scatter than observed for the density fluctuations. Large velocity fluctuations with 
$\Utilde \gtrsim 2.5$ are correlated with positive temperature fluctuations. Similar to the correlation
to density fluctuations, are large negative velocity fluctuations, $\Utilde \lesssim 2.5$ on average
in phase with small, positive temperature fluctuations.

%%%%%%%%%%%%%%%%%%%%%%%%%%%%%%%%%%%%%%%%%%%%%%%%%%%%%%%%%%%%%%%%%%%%%%%%%%%%%%%%%%%%%%%%%%%%%%%%%%%
% Created by mkplot_vrad_condavg_ne_te.py
% Figure 19
\begin{figure}
    \includegraphics{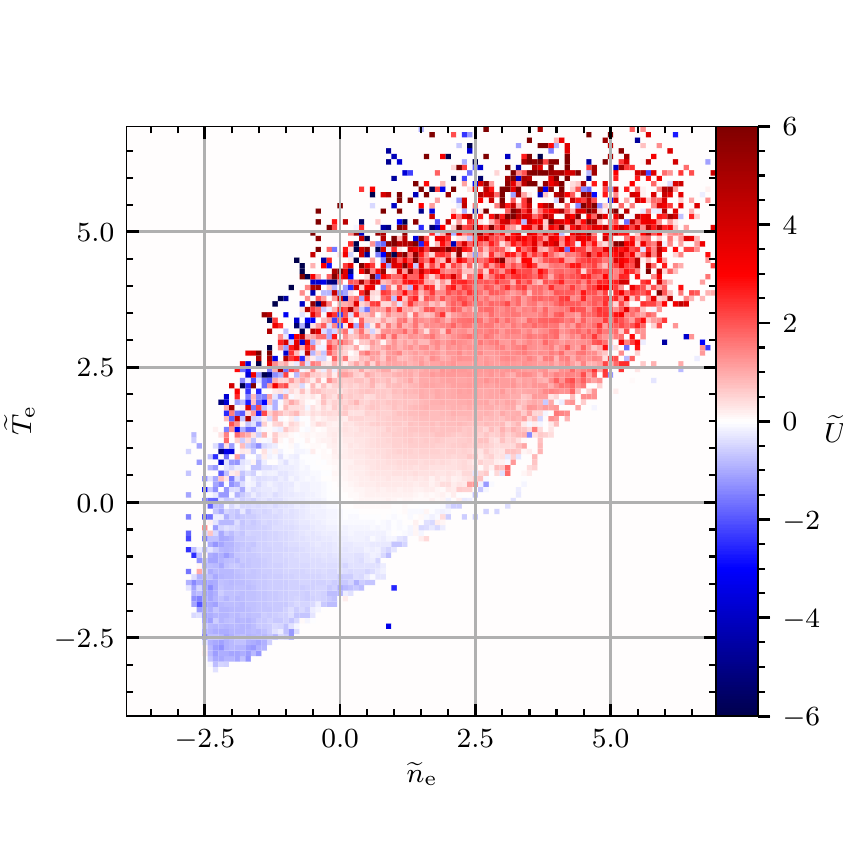}
    \caption{Amplitude of the velocity fluctuations as a function of the electron density 
             and temperature fluctuation amplitude.}  
    \label{fig:condavg2d_vrad}
\end{figure}
%%%%%%%%%%%%%%%%%%%%%%%%%%%%%%%%%%%%%%%%%%%%%%%%%%%%%%%%%%%%%%%%%%%%%%%%%%%%%%%%%%%%%%%%%%%%%%%%%%%

Figure \ref{fig:condavg2d_vrad} presents the radial velocity amplitudes encoded in a scatter
plot of the electron density and temperature fluctuations. Large-amplitude velocity
fluctuations are in phase with large-amplitude fluctuations in both electron temperature
and density. The magnitude of $\Utilde$ increases with the amplitude of
$\netilde$ and $\Tetilde$. Few negative velocity fluctuations are observed for
$\netilde \gtrsim 0$ and $\Tetilde \gtrsim 0$. Negative velocity fluctuations are
observed for $\netilde \gtrsim 0$ and $\Tetilde \lesssim 0$, as well as for
$\Tetilde \gtrsim 0$ and $\netilde \lesssim 0$.

\begin{figure}  
    \includegraphics{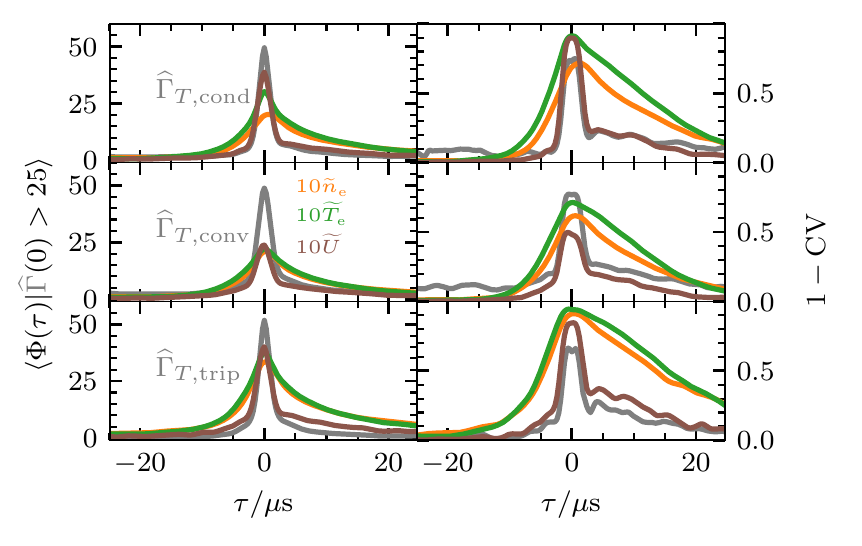}
    \caption{Conditionally averaged waveform of the electron density and temperature
             and the radial velocity during heat flux bursts (left panels). The right
             panels shows the conditional variance of the waveforms.}
    \label{fig:condavg_gammat}
\end{figure}

We continue by investigating how these fluctuations contribute to the radial heat flux.
For this we present the conditionally averaged wave forms of the fluctuating time series,
centered around heat flux events exceeding $25$ in normalized units, shown in
\Figref{condavg_gammat}. Here we use the conductive and convective heat flux, as well as
contributions from triple correlations as reference signals. The left panels show the
conditionally averaged waveforms of the respective heat fluxes and the right panels show
their conditional variance (CV) \Ref{oynes-1995}. The conditional variance describes the
average deviation of the individual waveforms from the average waveform. A value of  
$\mathrm{CV} = 0$ describes identical individual waveforms while a value of 
$\mathrm{CV} = 1$ describes random individual waveforms. In total $2692$ local maxima 
are identified in the conductive heat flux time series, $3963$ in the convective heat flux
time series and $992$ in the triple correlations time series. These counts agree with the
PDFs of individual heat flux contributions, shown in \Figref{histogram_gammat}, where the
same ordering of the sample probabilities holds for large amplitude fluctuations. 

Using the conductive heat flux as a reference signal we find that its auto-conditionally 
averaged waveform appears triangular and is reproducible with $\max 1 - \mathrm{CV}$ 
approximately $1$. The average waveform of the electric drift velocity appears similar 
in shape while the average waveform of the electron density and temperature fluctuations 
present a broad shape with a fast rise and slow decay. Temperature and velocity fluctuations 
which mediate conductive heat flux events appear highly reproducible. The slower decay time
scale of the temperature fluctuations mediating these heat flux events 
($\taud \approx 15 \mus$) appears reproducible among the individual events, as suggested by
the slowly decreasing conditional variance for $\tau > 0$.  
Density and velocity fluctuations, which constitute convective heat flux events present 
on average a qualitatively similar shape, although with an average amplitude smaller by a
factor of $1.5$. Their average waveform also appears more random than their average waveform
for conductive heat flux events. Here we find
$\max 1 - \mathrm{CV} \approx 0.6$ for $\netilde$ and 
$\max 1 - \mathrm{CV} \approx 0.7$ for $\Utilde$. As for the conductive heat flux, the
average density and temperature fluctuations associated with convective heat flux events
present less variance for $\tau > 0$ than for $\tau < 0$.
Conditionally averaged waveforms of the density, temperature, and velocity fluctuations
associated with large-amplitude triple correlation heat flux events are qualitatively
similar to the two previous cases. The average velocity waveform appears similar to
the average heat flux waveform. The average density and temperature waveform features
a fast rise and a slow decay. Their slower decay is robustly reproducible by all
individual density and temperature waveforms.

%%%%%%%%%%%%%%%%%%%%%%%%%%%%%%%%%%%%%%%%%%%%%%%%%%%%%%%%%%%%%%%%%%%%%%%%%%%%%%%%%%%%%%%%%%%%%%%%%%%
% Creatd by graphics/mkplot_condavg_gammat.py
% Figure 20

%%%%%%%%%%%%%%%%%%%%%%%%%%%%%%%%%%%%%%%%%%%%%%%%%%%%%%%%%%%%%%%%%%%%%%%%%%%%%%%%%%%%%%%%%%%%%%%%%%%

\section{Discussion} 
\label{sec:discussion}

The data time series of the ion saturation current, the electron density, and the electron 
temperature have similar statistical properties. They all feature large-amplitude, 
intermittent bursts on a similar time scale. However, the relative fluctuation level of the $\Is$ 
data is approximately twice as large as those of the $\nee$ and $\Te$ data. Furthermore the 
PDF of $\Istilde$ features higher probabilities for large-amplitude fluctuations than PDFs of 
$\netilde$ and $\Tetilde$. Sample skewness and flatness of the $\Istilde$ time series are 
approximately $1.5 - 2$ times larger than for the $\netilde$ and $\Tetilde$ time series. This
tendency of the $\Is$ data to deviate strongly from the sample mean may be explained by the
correlation of the $\nee$ and $\Te$ sample amplitude. According to \Eqnref{Isat} 
correlated large-amplitude $\nee$ and $\Te$ samples increase the ion saturation current significantly.

Relative fluctuation levels of the electron density and temperature are given by $0.27$
and $0.19$ respectively. Similar values have been reported from the TEXT-U tokamak
\Ref{meier-2001}. The ion saturation current samples feature a relative fluctuation
level of $0.47$. A lower relative fluctuation level of $\Te$ compared to $\Is$ was also
reported from ASDEX Upgrade \Ref{horacek-2010}. Coefficients of sample skewness
and excess kurtosis are given by $S=0.69$ and $F=0.79$ for $\netilde$ and by $S=0.63$ and
$F=0.88$ for $\Tetilde$. These values are similar in magnitude to values reported from the
SINP tokamak \Ref{saha-2008}. We note that the particular numerical value of the time series
statistics depends weakly on the width of the applied running average filter and weakly
on whether the time series are averaged over the MLP electrodes. However, the conclusion 
that electron density and temperature time series present large-amplitude, intermittent 
bursts, with skewed, non-Gaussian PDFs, is independent of the data preprocessing.

These observations motivate to interpret the $\Is$, $\nee$, and $\Te$ data time series as a 
realization of the stochastic process described by \Eqnref{shotnoise}. Indeed, the time
series are well described by a Gamma distribution. A least squares fit to the PDF predicted
by the stochastic model \Ref{theodorsen-2016-ppcf, theodorsen-2017-ps} yields shape 
parameters of $\gamma = 2.5$ for $\Istilde$ as well as $\gamma = 8.4$ and $\gamma = 8.5$ for 
$\netilde$ and $\Tetilde$ respectively. The shape parameter agrees well with results from
previous analysis of the ion saturation current in Alcator C-Mod, taken with conventional
Langmuir probes \Ref{kube-2016-ppcf}. On the other hand, PDFs of scrape-off layer 
fluctuations at the limiter radius, measured by the GPI diagnostic, are more skewed towards positive
values and present a shape parameters closer to unity \Ref{garcia-2013}. This disparity may be attributed
to the fact that GPI is sensitive to both the electron density and temperature. Furthermore
may burnouts, where hot blobs ionize neutral atoms, locally decrease the measured GPI
intensity level \Ref{zweben-2002, stotler-2003}.

Intermittent, large-amplitude fluctuations of the electron density and temperature may have
consequences for the life-time of the plasma facing components (PFCs) \Ref{marandet-2016}.
A Debye sheath at the vacuum vessel wall accelerates ions from the plasma onto the PFCs.
Assuming that the ion temperature is equal to the electron temperature, ions impact onto the
PFCs with an energy approximately five times the electron temperature \Ref{naujoks-book}.
Such processes are often quantified by the sputtering yield $Y$, which gives the ratio of
emitted target particle per incident. The sputtering yield is computed with the Bohdansky
formula \Ref{bohdansky-1984} which depends non-linearly on the impact particle energy and
on material properties of the target. Using the average temperature values reported here,
yields a vanishing sputtering yield, $Y(\mean{E}) = 0$. Using the fact that the impact
ions are Gamma distributed with a shape parameter $\gamma = 8.5$ we find 
$\mean{Y(E)} = 5.4 \times 10^{-5}$ for the Molybdenum walls installed in Alcator C-Mod. 
For lighter materials such as Beryllium, the used distribution results in average sputtering
yields larger by approximately three orders of magnitude.

The fluctuation amplitudes in the $\netilde$ and $\Tetilde$ time series, as well as the waiting
times between them are well described by an exponential distribution. The conditionally averaged
waveform of local maxima is well approximated by a two-sided exponential function with a duration
time of approximately $15\, \mus$. Similar conditionally averaged waveforms of electron density
and temperature fluctuations have been reported from DIII-D \Ref{boedo-2003, rudakov-2005}.
Measurements from ASDEX upgrade, suggesting a dip in the conditionally averaged temperature
waveform \Ref{horacek-2010}, are not confirmed by our analysis.

The power spectral densities of the $\netilde$ and $\Tetilde$ time series agrees well
with the PSD predicted by the stochastic model. Estimating the duration time by a 
least squares fit of \Eqnref{shotnoise_psd} to the data yields duration times of
approximately $15\, \mus$, comparable to the estimated duration time from fits to the
conditionally averaged waveform. On the other hand a fit to the PSD yields an asymmetry
parameter $\lambda = 0$ while conditional averaging yields $\lambda = 0.4$. This discrepancy
is likely due to significant overlap of the pulses in the data time series, as suggested
by $\gamma = \taud / \tauw \approx 10$ from \Figref{histogram_is_ne_te} \Ref{garcia-2012}.
Thus there are on average several pulses within a given conditional averaging window which
smear out the conditionally averaged waveform. 

The PDF of the $\Vptilde$ time series is skewed towards positive values. Given the sample correlation
coefficient of this time series and the $\Tetilde$ time series, $R_{\Vp, \Te} = 0.77$, this suggests
that the plasma potential is governed by the electron temperature. A comparison of velocity and flux
estimators using the density and plasma potential data to the ''classical'' ion saturation current
and floating potential data suggests however that this effect has no significant consequences. The
radial velocity estimates using either potential variable feature similar sample statistics, as shown
in \Figref{histogram_vrad}. Particle flux samples computed from $\nee$ and $\Vp$ data are almost
indistinguishable from samples computed using $\Is$ and $\Vf$ data, as shown in \Figref{histogram_gamman}.

The velocity time series has similarities to the electron density and temperature 
time series - it features intermittent, large-amplitude deviations from the sample mean.
Large amplitude deviations are however both positive and negative. The PDF of the time 
series is therefore symmetric. It also features exponential tails for large amplitude events
$| \Utilde | \gtrsim 2.5$. Generalizing the stochastic model to include Laplace 
distributed amplitudes yields an analytic expression for the PDF which describes 
the data time series over more than 3 orders of magnitude in probability. 
The average pulse duration in the $\Utilde$ data, $5\, \mus$, is three times smaller
than the pulse duration time in the electron density and temperature time series.
The shape parameter of the PDF is given by $10$, comparable to the shape parameter
that best describes the PDF of the $\netilde$ and $\Tetilde$ data.

A correlation analysis of the large-amplitude fluctuations in the electron density, temperature,
and velocity time series shows that a large fraction of them are in phase.
Density and velocity fluctuations that appear in phase lead to large particle flux events.
This may explain the elevated tail of the PDF of $\widehat{\Gamma}_n$, show in 
\Figref{histogram_gamman}. PDFs with similar shapes have been observed for scrape-off layer plasmas 
\Ref{xu-2005, garcia-2006-tcv, garcia-2007-nf, fedorczak-2011, kube-2016-ppcf}
as well as in numerical simulations of scrape-off layer plasmas 
\Ref{ghendrih-2003, ghendrih-2005, garcia-2008-ppcf}. Analysis of measurements taken in 
TEXT-U also report a strong correlation between density and temperature fluctuations \Ref{meier-2001}.
Figure \ref{fig:condavg2d_vrad} suggests a similar strong correlation between $\netilde$
and $\Tetilde$ in our time series. This figure furthermore suggests that a fraction of the 
large-amplitude fluctuations in all three quantities are correlated. These can be interpreted as 
dense and hot plasma blobs. 

This is compatible with the theory that the radial motion of plasma blobs is governed 
by the interchange mechanism. Fluid models, often employed to describe blob dynamics, 
suggest that the radial blob velocity is determined by the poloidal electron pressure 
gradient within the blob structure \Ref{krash-2001, bian-2003, garcia-2006}. 
Numerical simulations of such models find a dipolar potential structure aligned with the 
pressure gradient. While the dipole is out of phase with the pressure perturbation the 
resulting electric drift is in phase with the pressure perturbation. The resulting electric 
dipole structure has been reported from floating potential measurements in tokamak plasmas 
\Ref{rudakov-2002, grulke-2006, kube-2016-ppcf, theodorsen-2016-ppcf} 
and basic plasma experiments \Ref{katz-2008, furno-2011, theiler-2011}. 
Numerical simulations of blobs including dynamic finite Larmor radius effects present 
dipolar potential stratifications along the pressure gradient of a plasma blob 
\Ref{wiesenberger-2014, held-2016}. Furthermore numerical simulations of blobs, 
electrically connected to sheaths formed at the divertor, show a regime where a mono-polar
potential structure within a blob results in an intrinsic blob spin 
\Ref{myra-2004, dippolito-2004}. The observed mono-polar dip associated with large-amplitude
electron density maxima is thus inconclusive about whether a single physical mechanism
governs blob propagation.

The radial electron heat flux time series features a relative fluctuation level of approximately
three and non-gaussian statistics, skewed towards large-amplitude events. The conductive and
convective heat fluxes contribute $94\%$ to the total fluctuation driven heat flux. Furthermore is
their respective relative fluctuation level approximately the same as that of the total heat flux.
As shown in the joint PDFs of $\Utilde$, $\netilde$, and $\Tetilde$, see \Figsref{jointpdf_vrad_ne}
and \ref{fig:jointpdf_vrad_te}, and the conditional average analysis in \Figref{condavg_gammat},
is there slightly more scatter in the convective heat flux than in the conductive heat flux. Triple
correlations contributing on average approximately $6\%$ to the total heat flux. The large
relative fluctuation level of this contribution to the total flux, approximately $13$, is due to the
few number of events where large-amplitude fluctuations of $\Utilde$, $\netilde$, and $\Tetilde$ are
in phase, as shown in \Figsref{histogram_gammat} and \ref{fig:condavg2d_vrad}.

\section{Conclusions}
\label{sec:conclusion}
Plasma fluctuations in the out board mid-plane scrape-off layer plasma in an ohmically 
heated, lower single-null diverted discharge in Alcator C-Mod have been analyzed. One second
long data time series were sampled using Mirror Langmuir probes, dwelling at the outboard 
mid-plane limiter position. Time series of the electron density and temperature as well as
the ion saturation current present intermittent, large-amplitude bursts. Large-amplitude
fluctuations in the ion saturation current appear more frequently than similar 
large-amplitude fluctuations in the electron density and temperature time series.
Large-amplitude $\nee$ and $\Te$ fluctuations appear in phase. This leads to increased
$\Is$ samples compared to attributing them solely to electron density fluctuations.
Both $\netilde$ and $\Tetilde$ are shown to be well described by the stochastic process
given by \Eqnref{shotnoise}. We find furthermore that the velocity fluctuations can be
described by a similar stochastic process by allowing for both negative and positive
fluctuation amplitudes.
The particle and heat flux towards the outboard mid-plane limiter structure appear
intermittent and are driven by fluctuations in both the electron density and temperature.
Both conductive and convective heat feature a similar PDF and contribute respectively 
approximately $56$ and $38$ percent to the total heat flux. Hot and dense plasma blobs 
contribute to the heat flux via triple correlations, albeit on average approximately $6$ 
percent. Accounting for the observed fluctuations of the electron temperature shows
that large heat flux events contribute to sputtering of the plasma facing components.

Future work will focus on exploring the fluctuation statistics for various plasma 
parameters as well as analysis on fluctuations sampled by divertor probes. 

\section*{Acknowledgments}
This work was supported with financial subvention from the Research Council of Norway under
Grant No. 240510/ F20. Work partially supported by US DoE Cooperative agreement 
DE-FC02-99ER54512 at MIT using the Alcator C-Mod tokamak, a DoE Office of Science user 
facility. R.K., O.\ E.\ G.\, and A.\ T.\ acknowledge the generous hospitality of the MIT 
Plasma Science and Fusion Center.

\appendix
\label{sec:appendix}
\section{Stochastic Model}  
For an interpretation of the data time series we employ the stochastic model
developed in Refs. \Ref{garcia-2012, theodorsen-2016-ppcf, garcia-2016, garcia-2017-ac}.
Within this framework time series are modeled as the super-position of
uncorrelated pulses, 
\begin{align}
    \Phi(t) = \sum\limits_{k=1}^{K(T)} A_k \varphi \left( \frac{t - t_k}{\taud} \right). \label{eq:shotnoise}
\end{align}
Here $K(T)$ gives the number of pulses arriving in the time interval $[0:T]$,
$A_k$ gives the amplitude of the $k$-th pulse and $t_k$ its arrival time. A universal
pulse shape is given by $\varphi(\theta)$ and $\taud$ gives the characteristic time scale
of the pulses.

Motivated by measurements in scrape-off layer plasmas
\Ref{boedo-2003, antar-2003, garcia-2006-tcv, garcia-2007-nf, horacek-2010, garcia-2013, garcia-2015, militello-2013, carralero-2014}
and numerical simulations 
\Ref{garcia-2006-tcv, garcia-2007-nf} 
we assume that the pulse amplitudes are exponentially distributed and that all pulses
present the same pulse shape. We also assume that pulse arrivals are governed by a Poisson
process where $K$ pulses arrive in a time interval $[0:T]$ with an average waiting time $\tauw$.
The ratio of the pulses duration time and the average waiting time between pulses
$\gamma = \taud / \tauw$ is referred to as the intermittency parameter. Realizations of 
\Eqnref{shotnoise} with significant pulse overlap are described by large values of 
$\gamma$, while realizations of \Eqnref{shotnoise} with little pulse overlap are described 
by a small value of $\gamma$.

Based on the same observations we postulate that the average pulse shape is described by
a two-sided exponential function
\begin{align}
    \varphi(\tau) = 
        \begin{cases}
            \displaystyle{\exp \left(  \frac{\tau}{\taur} \right)} \quad & \mathrm{for}\, \tau < 0, \\
            \displaystyle{\exp \left( -\frac{\tau}{\tauf} \right)} \quad & \mathrm{for}\, \tau \geq 0. \\
        \end{cases}
    \label{eq:pulseshape}
\end{align}
The pulse duration time is given by the sum of the rise and fall e-folding times,
$\taud = \taur + \tauf$, and a pulse asymmetry parameter is defined as 
$\lambda = \taur / \taud$. Under these assumptions the process described by 
\Eqnref{shotnoise} is Gamma distributed \Ref{garcia-2012},
\begin{align}
    P_\Phi(\Phi) = \frac{1}{\mean{\Phi} \Gamma(\gamma)} \left(\frac{\Phi}{\mean{\Phi}} \right)^{\gamma - 1} \exp \left( -\frac{\Phi}{\mean{\Phi}}\right) \label{eq:PDF_gamma},
\end{align}
where $\mean{\cdot}$ denotes an ensemble average. The shape parameter of the PDF is given
by the intermittency parameter $\gamma$ and is notably independent of $\lambda$ and $\taud$.

The auto-correlation function of the normalized process 
$\widetilde{\Phi} = \left( \Phi - \mean{\Phi} \right) / \Phi_{\mathrm{rms}} $ is given by
\Ref{theodorsen-2016-ppcf, garcia-2016, garcia-2017-ac}
\begin{align}
    \mathcal{R}_{\widetilde{\Phi}}(\tau) = \frac{\tauf e^{-|\tau|/\tauf} - \taur e^{-|\tau|/\taur} }{\tauf - \taur}. \label{eq:shotnoise_acorr}
\end{align} 
This geometrical average approaches an exponential decay in the limit of large pulse
asymmetry, $\taur \ll \tauf$ or $\tauf \ll \taur$. For nearly symmetric pulses,
$\taur \approx \tauf$, the derivative of the auto-correlation function approaches zero for
small time lags, $\lim_{\tau \rightarrow 0^+} \mathcal{R}_{\widetilde{\Phi}}'(\tau) = 0$,
while $\mathcal{R}_{\widetilde{\Phi}}(t)$ decays exponentially for large time lags $\tau$.

Using \Eqnref{shotnoise_acorr}, one can show that the power spectral density of the 
process \Eqnref{shotnoise} is given by \Ref{garcia-2017-ac}
\begin{align}
    \mathrm{PSD}_{\widetilde{\Phi}}(\omega) = \frac{2 \taud}{\left[ 1 + \left( 1 - \lambda \right)^2 \taud^2 \omega^2 \right] \left[1 + \lambda^2 \taud^2 \omega^2 \right]}. \label{eq:shotnoise_psd}
\end{align}
This expression depends only on the pulse asymmetry parameter $\lambda$ and the duration
time  $\taud$ and is independent of the intermittency parameter $\gamma = \taud / \tauw$. 
For one-sided exponential pulses, $\lambda = 0$ or $\lambda = 1$ decays the PSD for large
frequencies $\omega$ as $\omega^{-2}$. Otherwise \Eqnref{shotnoise_psd} approaches 
$\omega^{-4}$ for large values of $\omega$.

%%%%%%%%%%%%%%%%%%%%%%%%%%%%%%%%%%%%%%%%%%%%%%%%%%%%%%%%%%%%%%%%%%%%%%%%%%%%%%%%%%%%%%%%%
\section{Assessing the impact of electron temperature outlier data points from MLP analysis}
Data time series, $\Is$, $\Vf$, and $\Te$, deduced from the MLP sometimes exhibit large peaks
which occur on time scales of approximately $1\, \mus$. These result in large values of sample
skewness and excess kurtosis, as listed in \Tabref{smoothing-stats}. Large-amplitude fluctuations
on this time scale are not observed by other diagnostics nor are they seen in numerical simulations.
We ascribe them to either uncertainties in the fit of the I-V characteristic performed by the MLP
analysis or to off-normal events, such as probe arcing.

Two approaches for identifying and treating the outliers in the MLP were performed. The first was
to smooth the current from the Langmuir electrode, sampled at the bias voltages $V^{+}$, $V^{-}$,
and $V^{0}$, using a running average filter. The filter window length used was $3$, $6$, $9$,
and $12$ points, corresponding to $0.9, 1.8, 2.7$ and $3.6\mus$. The difference between the raw
and the smoothed current time series gives an uncertainty on the input data for a fit to the I-V
characteristic. Table \ref{tab:smoothing-stats} lists the lowest order statistical moments of the
resulting fit parameter time series. The average value of both $\Is$ and $\Vf$ remains approximately
invariant when changing the length of the filter window. Their root mean square values vary only little
for filter radii larger than six. While their skewness and excess kurtosis significantly decreases with
increasing filter radius, they decrease little above a filter radius of 9 samples. The statistics of
the $\Te$ time series shows a slower convergence behavior. The time series average appears invariant
when applying the running average filter and its root mean square value changes only little. On the
other hand decrease the sample skewness and excess kurtosis significantly when applying the average
filter for filter radii less or equal six. Above this filter length these two sample coefficients
decrease only little.

\begin{table}[h!tb]
    \begin{tabular}{c|c|c|c|c}
                    & average                                   & rms                                       & S                                     & F \\ \hline
        $\Is$       & $18, 18, 18, 18, 18\, \mathrm{mA}$        & $6.4, 8.9, 8.7, 8.6, 8.4\, \mathrm{mA}$   & $2.3, 1.4, 1.3, 1.2, 1.2$             & $20, 4.9, 3.9, 2.4, 2.2$ \\ \hline
        $\Te$       & $13, 13, 13, 13, 13\, \mathrm{eV}$        & $5.3, 4.0, 3.5, 3.3, 3.2\, \mathrm{eV}$   & $8.1, 2.7, 1.3, 0.95, 0.71$           & $170, 26, 4.7, 2.2, 1.4$ \\ \hline
        $\Vf$       & $1.5, 1.5, 1.5, 1.5, 1.5\, \mathrm{V}$    & $6.0, 5.9, 5.7, 5.5, 5.3\, \mathrm{V}$    & $-0.89, -0.86, -0.84, -0.83, -0.82$   & $1.2, 1.0, 0.91, 0.92, 0.82$
    \end{tabular}
    \caption{Average, root-mean-square, skewness and excess kurtosis of the fit 
        data time series, where the fit input current samples were subject to a
        1, 3, 6, 9, and 12 point running average filter. Data is taken from the MLP
        at the southwest electrode.}
    \label{tab:smoothing-stats}
\end{table}

The second approach to treat outliers was to identify suspicious fits to the MLPs 
I-V characteristic. For this, the time series $\{\Te, (V^{+}-V^{-}) / \Te, \sigma_{\Te}\}$
from the four MLPs were combined into a 12-dimensional data time series. Here $\sigma_{\Te}$ gives
the uncertainty of the estimated $\Te$ parameter. An outlier in this data space may be a single
MLP reporting a significantly larger $\Te$ value than the other three MLPs, together with a large
uncertainty $\sigma_{\Te}$ and a smaller fit domain $(V^{+} - V^{-}) / \Te$. Such outliers in the
time series were detected using the \emph{isolation forest} algorithm
\Ref{liu-2012-iforest, scikit-learn}. The single input parameter for this algorithm is an a-priori
estimate of the fraction of outliers in the data sample. 
The next step is to reduce the detected outliers to data points where only a single $\Te$ 
sample deviates from the other three. For this, a two-sided Grubbs' test was performed on 
all four $\Te$ samples in each outlier \Ref{nist-handbook}. Finally an averaged $\Te$ time 
series was computed, ignoring single $\Te$ samples for which the Grubbs' statistic suggests 
it to be an outlier.

\begin{figure}
    \includegraphics[width=0.5\textwidth]{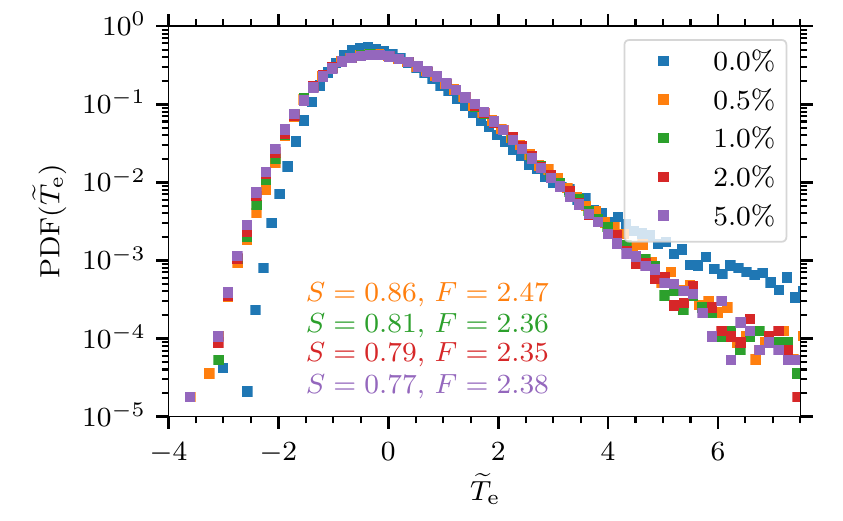}
    \caption{Histogram of the $\Tetilde$ time series subject to outlier removal, for
             using different a-priori assumption for the fractions of outliers.}
    \label{fig:hist_te_iforest}
\end{figure}

Figure \ref{fig:hist_te_iforest} compares histograms of $\Te$ samples, subject to the described
outlier removal process. The data denoted by $0.0\%$ is computed by averaging over all $\Te$ data
time series, including outliers. The other time series were calculated after removing outliers,
a-priori assuming $0.5 \ldots 5.0\%$ outliers. The input data for the I-V fits was smoothed using
a 3-point running average filter. Even assuming $0.5\%$ samples as outliers results in data with
significantly fewer large amplitude samples and with significantly smaller coefficients of sample
skewness and flatness. Further increasing the a-priori outlier fraction results in only minor
change of the data PDF and sample skewness and flatness.

%%%%%%%%%%%%%%%%%%%%%%%%%%%%%%%%%%%%%%%%%%%%%%%%%%%%%%%%%%%%%%%%%%%%%%%%%%%%%%%%%%%%%%%%%%%%%%%%%%%%
% Figure 21
%%%%%%%%%%%%%%%%%%%%%%%%%%%%%%%%%%%%%%%%%%%%%%%%%%%%%%%%%%%%%%%%%%%%%%%%%%%%%%%%%%%%%%%%%%%%%%%%%%%%

Even though this outlier removal procedure allows to regularize the $\Te$ data time series, does
it not allow to infer whether outlier $\Te$ samples are due to physical events, as large temperature
fluctuations, or nonphysical events, as probe arcing. The data analysis presented in this paper was
performed using fit parameter time series of $\Is$, $\Vf$, and $\Te$ subject to a 12 point Gaussian 
window. Choosing this window conserves the smoothing properties of the running-average window
of similar size, while at the same time it allows to avoid spurious oscillations in the high-frequency
power spectral density of the time series.

\bibliography{myrefs.bib}
\end{document}